\documentclass{aa}  

\usepackage{graphicx}
\usepackage{txfonts}
\usepackage{natbib}
\usepackage{amsmath} 
\usepackage{amssymb}
\usepackage{mathtools}
\usepackage{xcolor}
\usepackage{multirow}
\usepackage{array}
\usepackage{makecell}
\usepackage{siunitx}
\usepackage{ulem}
\usepackage{graphicx}
\usepackage{subcaption}

\usepackage[colorinlistoftodos]{todonotes}
\newcolumntype{P}[1]{>{\centering\arraybackslash}p{#1}}
\newcommand{\cm}[0]{CLM}
\newcommand{\ncm}[0]{NCLM}
\newcommand{\M}[1]{M#1}
\newcommand{\R}[1]{R#1}
\newcommand{\A}[1]{A#1}
\newcommand{\MS}[1]{MS#1}

\usepackage{xparse}

\ExplSyntaxOn
\NewDocumentCommand{\DefineDictionary}{mm}
 {
  \arclupus_dict_def:nn { #1 } { #2 }
 }

\seq_new:N \l__arclupus_dict_temp_seq

\cs_new_protected:Nn \arclupus_dict_def:nn
 {
  \prop_gclear_new:c { g_arclupus_#1_dict_prop }
  \clist_map_inline:nn { #2 }
   {
    \__arclupus_dict_add:nn { #1 } { ##1 }
   }
  \cs_new:cpn { #1 } ##1 { \prop_item:cn { g_arclupus_#1_dict_prop } { ##1 } }
 }

\cs_new_protected:Nn \__arclupus_dict_add:nn
 {
  \seq_set_split:Nnn \l__arclupus_dict_temp_seq { / } { #2 }
  \prop_gput:cxx { g_arclupus_#1_dict_prop }
   {
    \seq_item:Nn \l__arclupus_dict_temp_seq { 1 }
   }
   {
    \seq_item:Nn \l__arclupus_dict_temp_seq { 2 }
   }
 }
\cs_generate_variant:Nn \prop_gput:Nnn { cxx }
\ExplSyntaxOff

\DefineDictionary{zcluster}{m1206/0.439, m0416/0.397, r2248/0.346, m1149/0.542, r2129/0.234, m1931/0.352, m0329/0.450, m2129/0.587, m1115/0.352, a370/0.375, a2744/0.308, a383/0.188, m2137/0.316, m1311/0.494, r1347/0.451}

\newcommand{\angstrom}{\textup{\AA}}

\definecolor{agc}{rgb}{0.0, 0.26, 0.15}

\begin{document} 

   \title{The search for galaxy cluster members with deep learning of panchromatic HST imaging and extensive spectroscopy}


   \author{G. Angora\inst{1}\fnmsep\inst{3}\thanks{\email{gius.angora@gmail.com}},
           P. Rosati\inst{1}\fnmsep\inst{2}\fnmsep\inst{9},
           M. Brescia\inst{3},
           A. Mercurio\inst{3},
           C. Grillo\inst{4}\fnmsep\inst{5}\fnmsep\inst{6},
           G. Caminha\inst{7},
           M. Meneghetti\inst{2},
           M. Nonino\inst{8},
           E. Vanzella\inst{2},
           P. Bergamini\inst{2},
           A. Biviano\inst{8},
           \and
           M. Lombardi\inst{4}
          }

   \institute{Department of Physics and Earth Science of the University of Ferrara, Via Saragat 1, I-44122 Ferrara, Italy\and
   INAF - Astronomical Observatory of Bologna, via Gobetti 93/3, 40129 Bologna, Italy\and
   INAF - Astronomical Observatory of Capodimonte, Salita Moiariello 16, I-80131 Napoli, Italy\and
   Department of Physics of the University of Milano, via Celoria 16, I-20133 Milano, Italy\and
   Dark Cosmology Centre, Niels Bohr Institute, University of Copenhagen, Lyngbyvej 2, DK-2100 Copenhagen, Denmark\and
   INAF - IASF Milano, via A. Corti 12, I-20133 Milano, Italy \and
   Kapteyn Astronomical Institute, University of Groningen, Postbus 800, 9700 AV Groningen, The Netherlands\and
   INAF - Astronomical Observatory of Trieste, via G. B. Tiepolo 11, I-34131, Trieste, Italy\and INFN, Sezione di Ferrara, Via Saragat 1, 44122 Ferrara, Italy}

   \date{Received xxx; accepted xxx}


 
  \abstract
  {The next generation of extensive and data-intensive surveys are bound to produce a vast amount of data, which can be efficiently dealt with using machine-learning and deep-learning methods to explore possible correlations within the multi-dimensional parameter space.}
  {We explore the classification capabilities of convolution neural networks (CNNs) to identify galaxy cluster members (\cm{s}) by using Hubble Space Telescope (HST) images of fifteen galaxy clusters at redshift $0.19 \lesssim z \lesssim 0.60$, observed as part of the CLASH and Hubble Frontier Field programmes.} 
  {We used extensive spectroscopic information, based on the CLASH-VLT VIMOS programme combined with MUSE observations, to define the knowledge base. We performed various tests to quantify how well CNNs can identify cluster members on ht basis of imaging information only. Furthermore, we investigated the CNN capability to predict source memberships outside the training coverage, in particular, by identifying \cm{s} at the faint end of the magnitude distributions.} 
  {We find that the CNNs achieve a purity-completeness rate $\gtrsim90\%$, demonstrating stable behaviour across the luminosity and colour of cluster galaxies, along with a remarkable generalisation capability with respect to cluster redshifts. We concluded that if extensive spectroscopic information is available as a training base, the proposed approach is a valid alternative to catalogue-based methods because it has the advantage of avoiding photometric measurements, which are particularly challenging and time-consuming in crowded cluster cores. As a byproduct, we identified 372 photometric cluster members, with mag(F814)$<$25, to complete the sample of 812 spectroscopic members in four galaxy clusters RX~J2248-4431, MACS~J0416-2403, MACS~J1206-0847 and MACS~J1149+2223.}
  {When this technique is applied to the data that are expected to become available from forthcoming surveys, it will be an efficient tool for a variety of studies requiring \cm{} selection, such as galaxy number densities, luminosity functions, and lensing mass reconstruction.}

   \keywords{ galaxies: clusters: general, galaxies: photometry, galaxies: distances and redshifts, techniques: image processing, methods: data analysis }

   \titlerunning{Galaxy cluster membership with deep learning}
   \authorrunning{G. Angora, P. Rosati, M. Brescia, et al.}

   \maketitle
%
\section{Introduction}\label{sec:intro}
Over the past decade, the field of astrophysics has been experiencing a true paradigmatic shift, moving rapidly from relatively small data sets to the big data regime. Dedicated survey telescopes, both ground-based and space-borne, are set to routinely produce tens of terabytes of data of unprecedented quality and complexity on a daily basis. These volumes of data can be dealt with through a novel framework, delegating most of the work to automatic tools and by exploiting all advances in high-performance computing, machine learning, data science and visualisation \citep{brescia:2018b}. The paradigms of machine learning (ML) and deep learning (DL) paradigms embed the intrinsic data-driven learning capability to explore huge amounts of multi-dimensional data by searching for hidden correlations within the data parameter space. 

Here, we explore the application of ML techniques in the context of studies of galaxy clusters, more specifically, to identify cluster members (\cm{s}) based on imaging data alone. In fact, obtaining a highly complete sample of spectroscopic members is an extremely expensive and time-consuming task, which can be simplified and accelerated thanks to the use of a limited amount of spectroscopic information in training ML methods.

Disentangling \cm{s} from background and foreground sources is an essential step in the measurement of physical properties of galaxy clusters, measuring, for example, the galaxy luminosity and stellar mass functions (e.g. \citealt{annunziatella2016, annunziatella2017}), in addition to studies of the cluster mass distribution via strong and weak lensing techniques (e.g. \citealt{caminha2017b, caminha2019, lagattuta2017, medezinski2016}). In particular, the study of the inner mass substructure of cluster cores with high-precision strong-lensing models and their comparison with cosmological simulations requires the simultaneous identification of background multiply lensed images and member galaxies to separate the sub-halo population from the cluster projected total mass distribution (e.g. \citealt{grillo2015, bergamini2019}). Such studies provide tests for structure-formation models and the cold dark matter paradigm \citep{diemand2011, meneghetti2020}. The need for efficient and reliable methods to identify cluster member galaxies from the overwhelming population of fore and background galaxies will become particularly pressing when a vast amount of photometric information becomes available with forthcoming surveys with, for example, the Large Synoptic Survey Telescope \citep[LSST, ][]{LSST2019} and Euclid \citep{Euclid2014}.

Owing to their ability to extract information from images, convolution neural networks \citep[CNNs,][]{LeCun:1989} have been widely used in several astrophysical applications, generally showing higher robustness and efficiency with respect to traditional statistical approaches. For example, they have been applied to phase-space studies of mock distributions of line-of-sight velocities of member galaxies at different projected radial distances. These DL techniques were able to reduce the scatter of the relation between cluster mass and cluster velocity dispersion by $\sim35$\% and by $\sim~20$\% when compared to similar ML methods, for instance, the support distribution machines \citep{Ho2019}. A similar DL approach has been successfully used to predict cluster masses from mock Chandra X-ray images, by limiting the parameter space to photometric features only, thus minimising both bias ($\sim5$\%) and scatter ($\sim12$\%), on average \citep{Ntampaka2015,Ntampaka2016,Ntampaka2019}. Such CNNs were also successfully used to discriminate between degenerate cosmologies, including modified gravity and massive neutrinos, by inspecting simulated cluster mass maps. \cite{Merten2019} showed that the DL techniques are able to capture distinctive features in maps mimicking lensing observables, improving the classification success rate with respect to classical estimators and map descriptors.

In recent years, the selection of \cm{s} has been addressed in several ways: via the classical identification of the members' red-sequence in colour-magnitude diagrams, aided by spectroscopic measurements (e.g. \citealt{caminha2019} for strong lensing applications); by measuring photometric redshifts with a Bayesian method \citep{molino2017, molino2019}; by exploiting an ML approach based on the so-called multi-layer perceptron trained by a quasi-Newton approximation \citep{Biviano2013,cav2015,bre2013}; or by fitting a multivariate normal distribution to the colour distribution of both spectroscopic members and field galaxies \citep{grillo2015}. All these methods require accurate photometric measurements, which are difficult to obtain with standard photometric techniques in galaxy clusters, due to the strong contamination from bright cluster galaxies, including the brightest cluster galaxies (BSGs), and the intra-cluster light \citep{molino2017}.

In this work, we exploit the paradigm of DL by designing a CNN that is able to identify cluster members using only HST images, based on CLASH \citep{postman2012} and Hubble Frontier Fields \citep[HFF, ][Koekemoer et al. in prep.]{Lotz2017} surveys and spectroscopic observations for the training set obtained with the VIMOS and MUSE spectrographs at the VLT.

The paper is structured as follows. In Sect.~\ref{sec:data}, we describe the HST imaging, spectroscopy measurements, and data configuration. We introduce the adopted DL approach in Sect.~\ref{sec:method}, including a synthetic description of the training setup and the metrics used to evaluate the network performance. In Sect.~\ref{sec:exps}, we illustrate details regarding the experiment configuration and results, as well as presenting a comparison of our model capabilities with other methods. In Sect.~\ref{sec:run}, we describe the process to identify new members by complementing the spectroscopic catalogues. We discuss in Sect.~\ref{sec:discussion} the potential and limitations of the method. Finally, we draw our conclusions in Sect.~\ref{sec:conc}. 

Throughout the paper, we adopt a flat $\Lambda$CDM cosmology model with $\Omega_M$=0.3, $\Omega_\Lambda$= 0.7, and H$_0$=70 km s$^{-1}$Mpc$^{-1}$. All of the astronomical images are oriented with north at the top and east to the left. Unless otherwise specified, magnitudes are in the AB system.

\section{Data layout}\label{sec:data}
   \begin{figure*}[htbp]
   \centering
   \includegraphics[width=2\columnwidth]{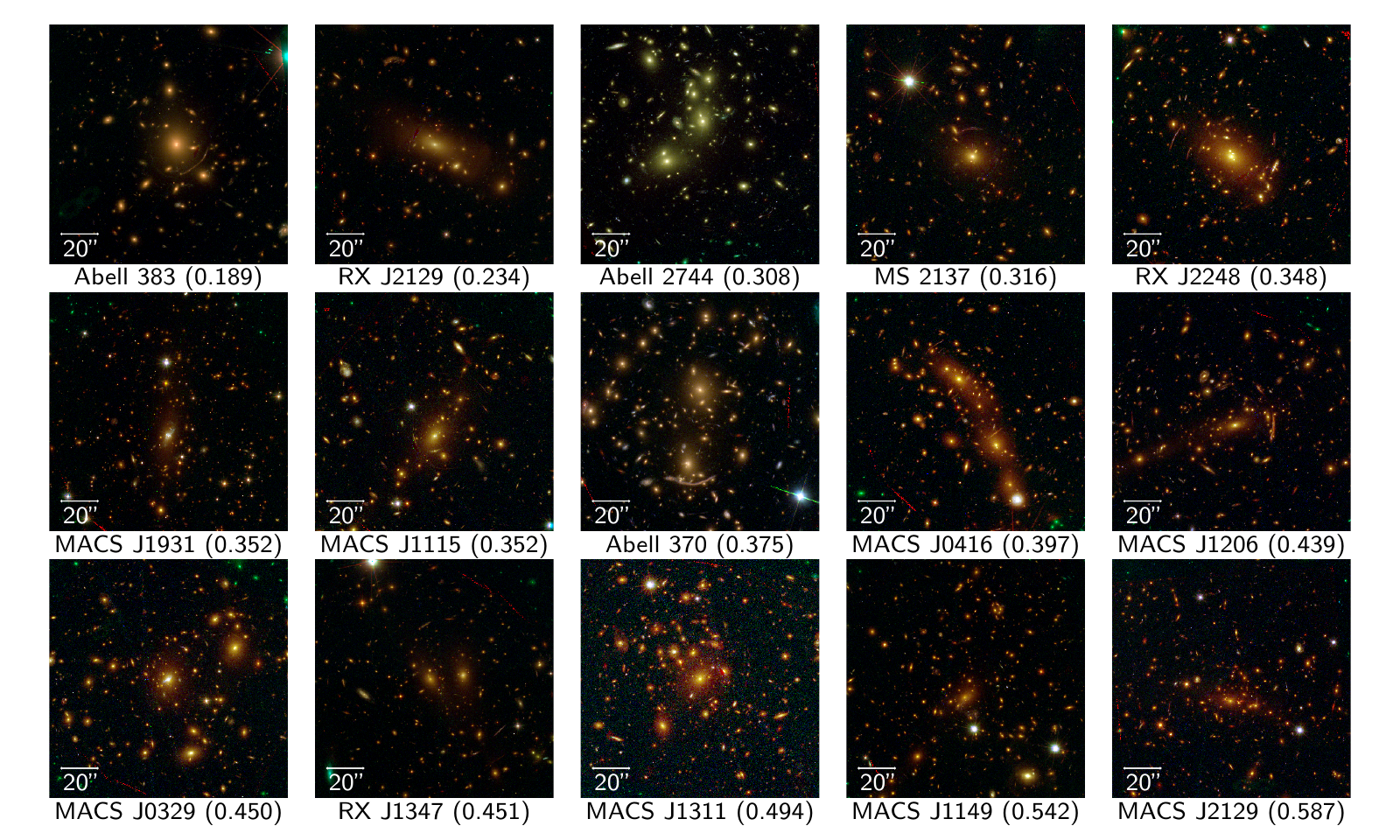}
\caption{Colour-composite images of the $15$ clusters included in our analysis, obtained by combining HST bands from optical to near IR. The images are squared cut-outs, $\sim 130\arcsec$ across, centred on the cluster core.}\label{fig:clusters}
   \end{figure*}

   \begin{figure}[htbp]
   \centering
   \includegraphics[width=0.95\columnwidth]{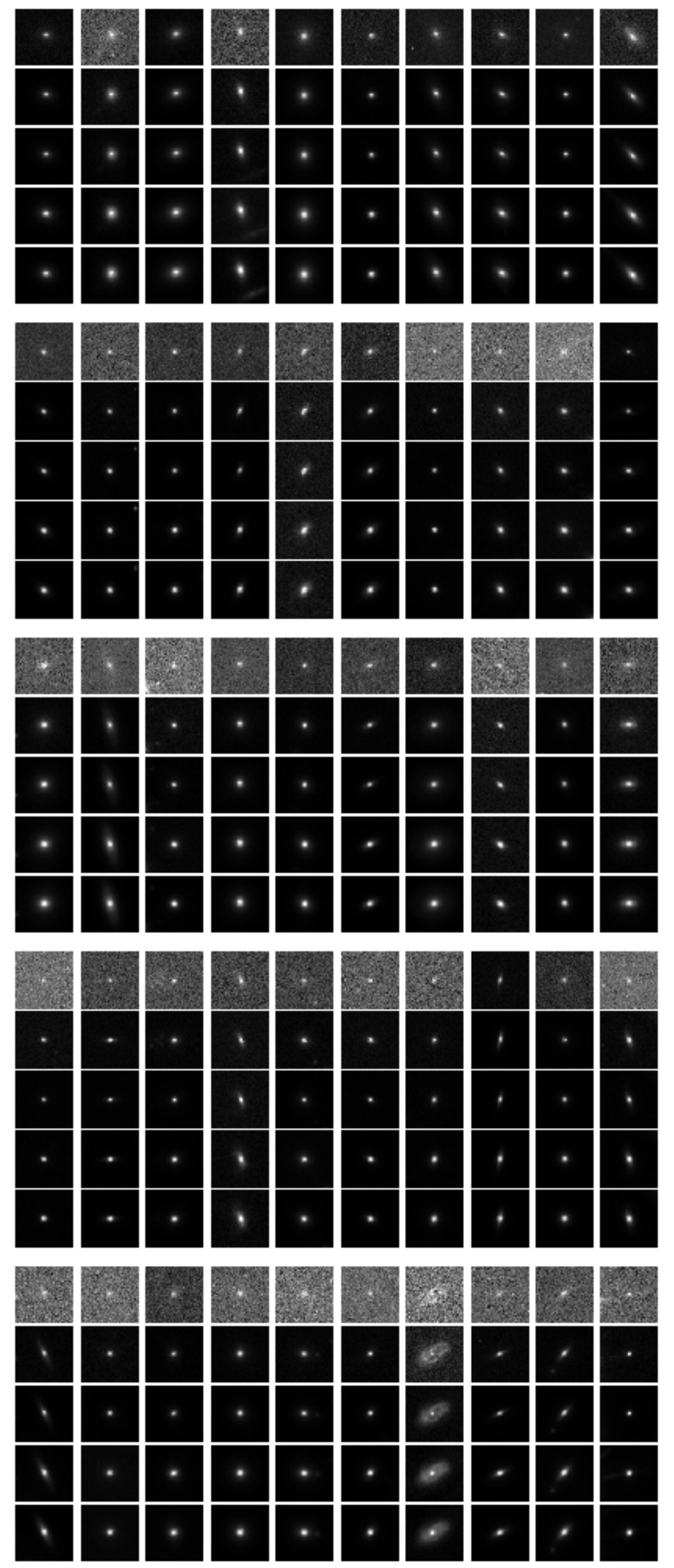}
   \caption{Examples of cut-outs of cluster members extracted from HST images ($F435$, $F606$, $F814$, $F105$, $F140$ bands), corresponding to five clusters (from top to bottom): \A{383} ($z=\zcluster{a383}$), \R{2248} ($z=\zcluster{r2248}$), \M{0416} ($z=\zcluster{m0416}$), \M{1206} ($z=\zcluster{m1206}$), \M{1149} ($z=\zcluster{m1149}$). All the cut-outs are 4 arcsec across.}\label{fig:cutouts}
   \end{figure}

In order to build a knowledge base, that is, to label a set of sources deemed suitable for training the neural network, we used the spectroscopic information based on the CLASH-VLT VIMOS programme (ESO 200h Large Program 186.A-0798, "Dark Matter Mass Distributions of Hubble Treasury Clusters and the Foundations of $\Lambda$CDM Structure Formation Models", PI: P. Rosati; \citealt{Rosati2014}), combined with archival observations carried out with the MUSE spectrograph \citep{bacon2014} (see  Table~\ref{tab:datanumbers}).

In the spectroscopic catalogues, we defined the \cm{s} as those having velocities $|v|\leq 3000$ kms$^{-1}$, with respect to the cluster rest-frame central velocity \citep{grillo2015, caminha2016, caminha2017a}. On the contrary, non-cluster-members (\ncm{s}) were those having greater differences in velocity. 

Cluster images were acquired by the HST ACS and WFC3 cameras as part of the CLASH \citep{postman2012} and HFF \citep{Lotz2017} surveys. The images were calibrated, reduced and then combined into mosaics with spatial resolutions of $0.065\arcsec$ \citep[see][]{Koekemoer2007, Koekemoer2011}. The fifteen clusters used in our study are shown in Fig.~\ref{fig:clusters}. Colour images were produced with the \textit{Trilogy} code \citep{Coe2012}, by combining HST filters from the optical to the near-infrared (NIR). Among the $16$ available HST filters used in our experiments, we considered bands covering the spectral range $4000 \angstrom - 16000\angstrom$ \citep{postman2012}, that is, the optical and NIR bands, excluding the UV filters for which the signal-to-noise ratio (S/N) of faint \cm{s} was too low.

For each spectroscopic source within the HST images, we extracted a squared cut-out with a side of $\sim 4\arcsec$ ($64$ pixels), centered on the source position. A sample of the dataset is shown in Fig.~\ref{fig:cutouts}, where \cm{s} were extracted from five clusters:  Abell~383 (\A{383}, $z=\zcluster{a383}$), RX~J2248-4431\footnote{Also known as Abell~S1063} (\R{2248}, $z=\zcluster{r2248}$), MACS~J0416-2403 (\M{0416}, $z=\zcluster{m0416}$), MACS~J1206-0847 (\M{1206}, $z=\zcluster{m1206}$), and MACS~J1149+2223 (\M{1149}, $z=\zcluster{m1149}$). Due to different pointing strategies and the fields of view of HST cameras, many sources do not have a complete photometric coverage, especially in the IR range. As a result, these objects with missing information were not useful for the training process \citep{BatistaMonard2003,marlin2008,parker2010}. With the aim of maximising the number of training samples with available spectroscopic redshift information, we chose four different band configurations:
\begin{itemize}
    \item[-] \textit{ACS}: only the seven optical bands (i.e. $F435$, $F475$, $F606$, $F625$, $F775$, $F814$, $F850$) were included in the training set, obtaining $1603$ \cm{s} and $1899$ \ncm{s};
    \item[-] \textit{ALL}: the training set involved all twelve bands (i.e. the seven optical bands and the five IR bands $F105$, $F110$, $F125$, $F140$, $F160$), thus reducing the number of objects to $1156$ and $1425$, respectively for \cm{s} and \ncm{s}, due to the rejection of missing data;
    \item[-] \textit{Mixed}: we selected five bands, corresponding to the filters available in the Hubble Frontier Fields survey, covering the optical-IR range, namely, $F435$, $F606$, $F814$, $F105$, $F140$, respectively. This includes $1249$ \cm{s} and $1571$ \ncm{s}; 
    \item[-] \textit{Mixed*}: same band combination as in the previous case (\textit{mixed}), but including two further clusters, namely, Abell~2744 (\A{2744}) and Abell~370 (\A{370}), for which only HFF imaging were available. This set is composed of $1629$ \cm{s} and $2161$ \ncm{s}.
\end{itemize}
In practice, the three configurations, \textit{ACS}, \textit{ALL} and \textit{mixed}, share the same clusters, while exploring different spectral information by varying the number of  sources. The \textit{mixed*} configuration considers an augmented cluster data set by including additional spectroscopic members.  A summary of the cluster sample and the spectroscopic data sets is given in Tab.~\ref{tab:datanumbers}.

\begin{table*}[htbp]
\caption{Cluster sample description.}
\label{tab:datanumbers}
\resizebox{2\columnwidth}{!}{
\begin{tabular}{llcccccccccc}
\hline
&&&&&\multicolumn{2}{c}{\textit{mixed*} (\textit{mixed})}&\multicolumn{2}{c}{\textit{ACS}}&\multicolumn{2}{c}{\textit{ALL}}\\\hline
Cluster & & $z_{cluster}$ & $z_{min}$ & $z_{max}$ & \cm{s} & \ncm{s} & \cm{s} & \ncm{s} & \cm{s} & \ncm{s} & ref\\\hline
\object{Abell~383}  & \A{383} & 0.188 & 0.176 & 0.200 & 59 & 51 & 91 & 79 & 59 & 51 & (1, 2)\\
\object{RX~J2129+0005} & \R{2129} & 0.234 & 0.222 & 0.246 & 47 & 124 & 66 & 132 & 40 & 118 & (3, 1)\\
\object{Abell~2744} & \A{2744} & 0.308 & 0.288 & 0.331 & 156$^{(a)}$ & 279$^{(a)}$ & \multicolumn{4}{c}{only frontier-field bands} & (4, 1)\\
\object{MS~2137-2353} & \MS{2137} & 0.316 & 0.303 & 0.329 & 45 & 49 & 70 & 80 & 45 & 49 & (3, 1)\\
\object{RX~J2248-4431}$^{(b)}$ & \R{2248} & 0.346 & 0.332 & 0.359 & 131 & 112 & 203 & 166 & 117 & 86 & (5, 1) \\
\object{MACS~J1931-2635} & \M{1931} & 0.352 & 0.338 & 0.365 & 68 & 97 & 80 & 110 & 65 & 96 & (3, 1)\\
\object{MACS~1115+0129} & \M{1115} & 0.352 & 0.338 & 0.365 & 78 & 69 & 116 & 111 & 62 & 55 & (3, 1)\\
\object{Abell~370}  & \A{370} & 0.375 & 0.361 & 0.389 & 224$^{(a)}$ & 311$^{(a)}$ & \multicolumn{4}{c}{only frontier-field bands} & (6, 1)\\
\object{MACS~J0416-2403} & \M{0416} & 0.397 & 0.382 & 0.410 & 237 & 277 & 266 & 287 & 227 & 230 & (7, 8, 9, 1)\\
\object{MACS~J1206-0847} & \M{1206} & 0.439 & 0.425 & 0.454 & 172 & 216 & 226 & 242 & 149 & 203 & (10, 1)\\
\object{MACS~J0329-0211} & \M{0329} & 0.450 & 0.435 & 0.464 & 74 & 76 & 104 & 104 & 66 & 73 & (3, 1)\\
\object{RX~J1347-1145} & \R{1347} & 0.451 & 0.438 & 0.467 & 56 & 107 & 71 & 120 & 56 & 107 & (3, 1)\\
\object{MACS~J1311-0310} & \M{1311} & 0.494 & 0.477 & 0.507 & 52 & 54 & 69 & 95 & 52 & 54 & (3, 1)\\
\object{MACS~J1149+2223} & \M{1149} & 0.542 & 0.527 & 0.558 & 141 & 237 & 149 & 270 & 129 & 202 & (11, 12, 1)\\
\object{MACS~J2129-0741} & \M{2129} & 0.587 & 0.571 & 0.603 & 89 & 102 & 92 & 103 & 89 & 101 & (1, 3)\\
\hline
\multicolumn{5}{c}{TOTAL}& 1629 & 2161 & 1603 & 1899 & 1156 & 1425 \\
\multicolumn{5}{c}{} & (1249) & (1571) & & & &  \\
\hline
\end{tabular}
}
\tablefoot{The name of the clusters, their redshift and their spectroscopic range to identify \cm{s} are reported in the first $5$ columns. The four band configurations, described in Sect.~\ref{sec:data}, are listed in columns $6$ to $11$. The references for each cluster can be found in the last column.\\
$^{(a)}$  Different spectroscopic data sets are described in the text. The case \textit{mixed} is similar to the \textit{mixed*} one, with the only difference that it does not include the two clusters \A{2744} and \A{370}.\\
$^{(b)}$ The cluster RX~J2248.7$-$4431 is also known as Abell~S1063.}
\tablebib{
(1)~\cite{Rosati2020}; (2)~\cite{monna2015}; (3)~\cite{caminha2019}; (4)~\cite{mahler2018};
(5)~\cite{caminha2016};(6)~\cite{lagattuta2019}; (7)~\cite{grillo2015}, (8)~\cite{balestra2016}; 
(9)~\cite{caminha2017a}; (10)~\cite{caminha2017b}; (11)~\cite{grillo2016}; (12)~\cite{treu2016}.
}

\end{table*}

\section{Methodology}\label{sec:method}
In this work, we discuss the results achieved by a VGGNET-like\footnote{We tested different network architectures, e.g. Residual~Net~X \citep{he2015, Xie:2016} and Inception Net \citep{szegedy:2014}. Due to their lower performances, we limited the description of the results to the VGGNET-like model, to avoid weighing down the text.} model, which is a CNN implementation inspired by the VGG network proposed by \cite{simonyan:2014}. 

As is customary in applications of ML methods, the data require a preparation phase, which, in this case, consisted of a data augmentation procedure, that is meant to construct a consistent labelled sample, followed by a partitioning of the dataset into training, validation, and blind testing sets.

Regarding data augmentation, given the relatively small sample of spectroscopic sources with respect to the typical size of the knowledge base required by supervised ML experiments, we increased the training set by adding images of spectroscopic sources, obtained from the original ones, through rotations and flips. The inclusion of these images in the training set also offered the possibility to make the network invariant to these operations, which works as an advantage for astronomical images as there is no defined orientation for the observed sources.

Concerning the partitioning of the data set, in order to fully cover the input parameter space, we opted for a stratified k-fold partitioning approach \citep{hastie2009, Kohavi1995}: the whole data set was split into $k=10$ non-overlapping folds, of which, iteratively, one extracted subset was used as a blind test set, while the others were taken as a training set. Such an approach has several advantages: \textit{(i)} increase of the statistical significance of the test set; \textit{(ii)} the blind test is performed only on original images; and \textit{(iii)} complete coverage of both training and test sets, keeping them well-separated at the same time.

The classification performance, obtained through all the experiments performed by this procedure, was evaluated by adopting a set of statistical estimators, directly derived from the classification confusion matrix \citep{stehman:1997}, namely, the classification efficiency (AE), averaged over the two classes (members and non-members), the purity (pur), the completeness (comp), and the harmonic mean of purity and completeness (F1, see \ref{ss:estimators}). The last three estimators have been measured for each class. 
Completeness (also known as recall) and purity (also known as precision) are the most interesting estimators, suitable for measuring the quality of the classification performed by any method. The completeness, in fact, measures the capability to extract a `complete' set of candidates of a given class, while purity estimates the capability of selecting a `pure' set of candidates (thus, minimising the contamination). Therefore, the classification quality is usually based on either one of such two estimators or their combination, depending on the specific interest of an experiment \citep{D'Isanto20163119}. In our case, we were most interested in finding the best trade-off between both estimators for the cluster members. 
The statistical evaluation was completed by also using the receiver operating characteristic curve \citep[ROC,][]{hanley:1982}, which is a diagram where the true positive rate (TPR, i.e. the completeness rate) is plotted versus the false positive rate (FPR, i.e. the contamination rate, which corresponds to $1-$ purity) by varying the membership probability threshold. The model performances are measured in terms of the area under the curve (AUC), thus providing an aggregate measure of performance across all possible classification thresholds.

A full description of the data preparation procedure and the statistical estimators is given in Appendix~\ref{app:method}, while details about the architecture and configuration of the DL model are reported in Appendix~\ref{app:CNN}.

\section{Experiments}\label{sec:exps}
In this section, we describe several experiments designed to test the performance of the CNNs and other methods. Specifically, with the data described in Sect.~\ref{sec:data}, we performed the following tests or experiments:
\begin{itemize}
    \item[-] \textit{EXP1}: efficiency of the DL approach by stacking the data of all the clusters in terms of:
    \begin{itemize}
        \item[-] \textit{EXP1a}: global evaluation
        \item[-] \textit{EXP1b}: redshift-dependence, namely separating \cm{s} into redshift bins;
    \end{itemize}
    \item[-] \textit{EXP2}: magnitude or colour dependence, by stacking data of a group of three clusters and varying their redshift range through:
    \begin{itemize}
        \item[-] \textit{EXP2a}: separating bright and faint sources
        \item[-] \textit{EXP2b}: separating red and blue galaxies
    \end{itemize}
    \item[-] \textit{EXP3}: a comparison of performances of our image-based CNN technique with other approaches, based on photometric measurements of field and cluster galaxies
\end{itemize}

\begin{figure}[htbp]
   \centering
   \includegraphics[width=\columnwidth]{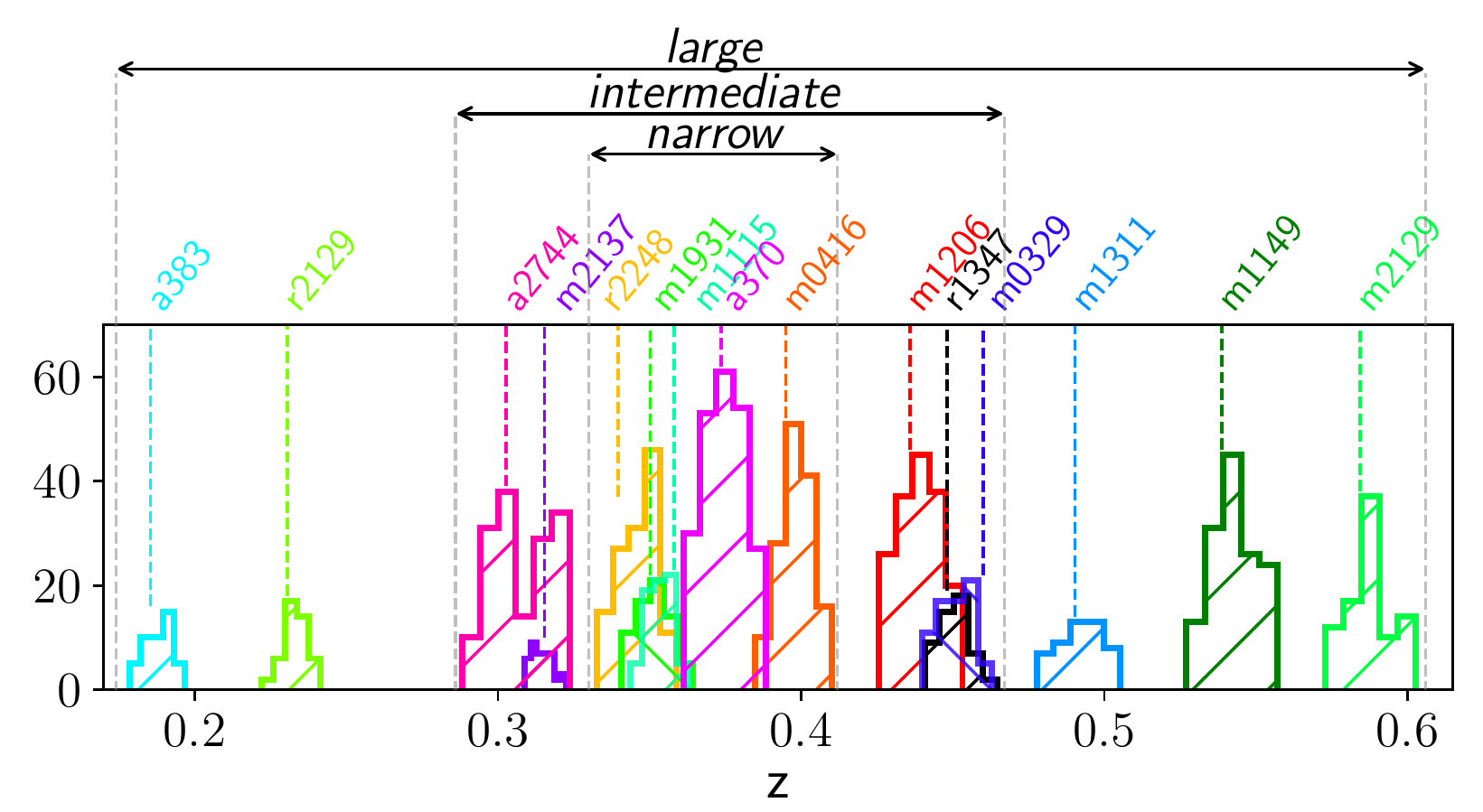}
   \caption{Redshift distribution of $1629$ spectroscopic members used for the \textit{EXP2} configuration. The three clusters \A{370} ($z=\zcluster{a370}$, 224 \cm{s}), \MS{2137} ($z=\zcluster{m2137}$, 45 \cm{s}) and \M{0329} ($z=\zcluster{m0329}$, 74 \cm{s}) are used as blind test set.}\label{fig:exp2:layout}
\end{figure}

\subsection{\textit{EXP1}: Combination of all clusters}\label{ss:exp1}

\begin{figure}[tbp]
   \centering
   \includegraphics[width=\columnwidth]{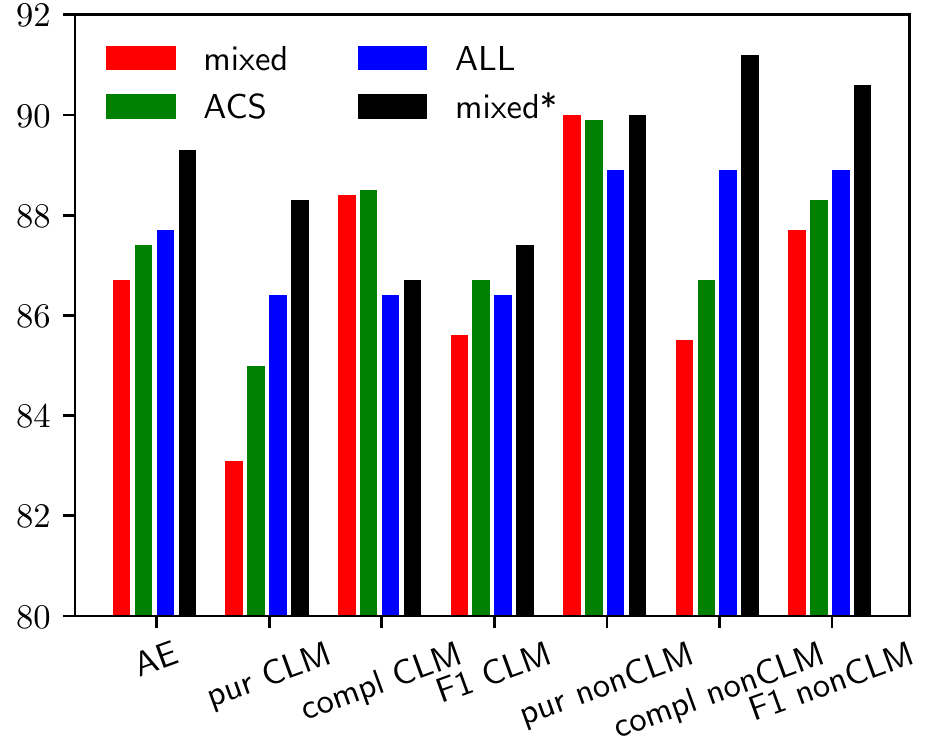}
   \caption{Performance percentages of the CNN in the \textit{EXP1} experiment with the four band configurations (see Sect.~\ref{sec:data}) in terms of the statistical estimators described in Sect.~\ref{ss:estimators}.}\label{fig:exp1:stack}
\end{figure}

At the first stage, we evaluated the global efficiency of a DL approach including all the available clusters, regardless of their redshift (ranging between $0.2$ and $0.6$), by exploring different combinations of photometric bands (as described in Sect.~\ref{sec:data}) and assembling the data set by stacking the information from all the images extracted from our cluster sample. We wanted to verify that DL models, given their intrinsic generalisation capabilities, were able to learn how to disentangle cluster members from non-member (foreground or background) sources, independently from the cluster redshift (\textit{EXP1a}). This although their members have different characteristics, such as apparent magnitudes or sizes, and also different signal-to-noise ratio at a fixed apparent magnitude, due to the different image depths. The results are shown in Fig.~\ref{fig:exp1:stack} and Table~\ref{tab:exp1:stack}, as a function of the band configuration, described in Sect.~\ref{sec:data}. 

For \ncm{}, we found similar values of the average efficiency ($87\%-89\%$), the purity (stable around $\sim90\%$) and the F1-score (with variations within $1.5\%$), regardless of band configuration. On the other hand, the \cm{} identification was, in general, characterised by larger variation ($83\%-91\%$) in the statistical estimators. With the \textit{mixed*} configuration, CNN achieved the best performances for \cm{} and it was also very stable in terms of \ncm, reaching an overall efficiency of $\sim89\%$. 

We also show, in Appendix~\ref{app:Exp1}, the estimators obtained for each cluster (Table~\ref{tab:exp1:clusters} and Fig.~\ref{fig:exp1:sizedep}). This analysis confirmed that the \textit{mixed*} combination showed the highest statistical values for all the thirteen clusters. Moreover, as expected, we demonstrated that there is a clear improvement of classification capabilities as the number of sources increases (an accuracy gain of $\sim2.3\%$ for an increment of $500$ samples). Furthermore, fluctuations of these estimators tend to be better constrained for a large set of objects, stabilising around $3\%$ when the number of samples is $\ge2000$ and showing an average reduction of $\sim9\%$ by quadrupling the number of sources.

\begin{figure}[htbp]
   \centering
   \includegraphics[width=\columnwidth]{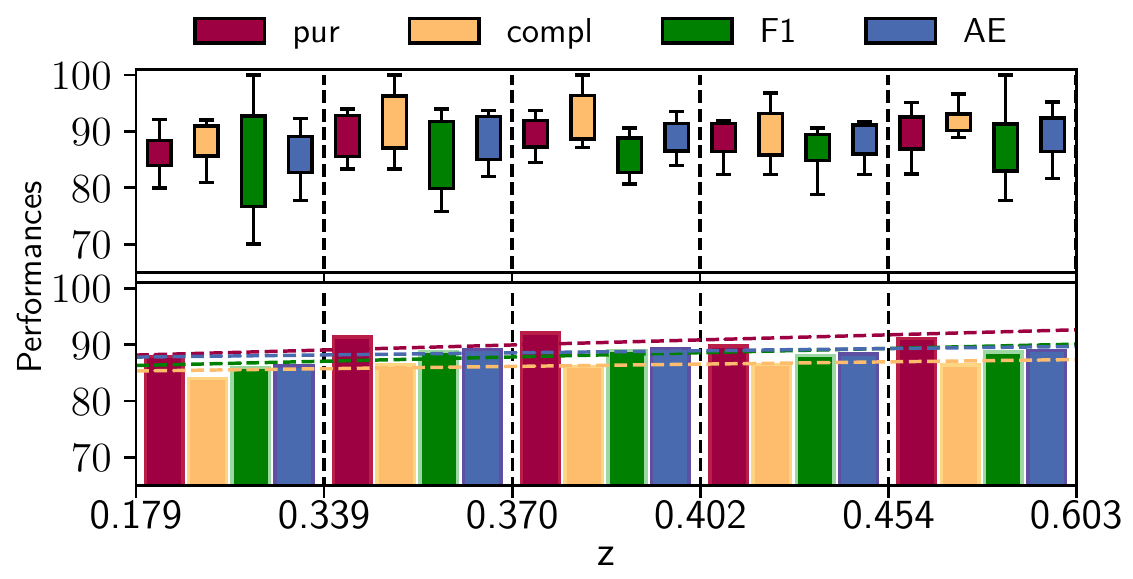}
   \caption{Percentages of CNN classification results for the four statistical estimators, measured as a function of \cm{} redshift range (\textit{EXP1}). The top panel describes their fluctuation in each bin, with the boxes delimiting the $25$th and $75$th percentiles (first and third quartile) and error bars enclosing the maximum point variations. The bottom panel shows the same metrics globally evaluated in each redshift bin, together with the best-fit lines.}\label{fig:exp1:zbin}
\end{figure}

Since the training set we used in this study was composed of galaxies spanning a large redshift range, as part of \textit{EXP1}, we investigated whether any dependence on redshift is present. To this aim, the \cm{} redshift range was split into five equal-sized bins ($\sim280$ samples). The performances and fluctuations related to the \textit{mixed*} band combination are shown in Fig.~\ref{fig:exp1:zbin}, while details on the metrics are given in Table~\ref{tab:exp1:zbin}. Despite the dissimilarities between galaxies at different depths, the CNN did not seem to be affected by the \cm{} redshift. In fact, CNN performances achieved in different redshift bins were all comparable, with a dispersion included within $0.04-1\sigma$ for the $65\%$ of cross-compared estimator pairs and a mean separation of $\sim0.8\sigma$.

Since the \textit{mixed*} band combinations provided the best results, all further experiments in the next sections refer to this band configuration.

\subsection{\textit{EXP2}: Selection of clusters as blind test set}\label{ss:exp2}

A second set of experiments was devoted to the study of the CNN capability to predict cluster membership of sources belonging to clusters that are not included in the training set, that is, avoiding having member galaxies belonging to the same cluster populating both the training and test sets. Thus, we considered \A{370} ($z=\zcluster{a370}$), MS~2137-2353 (\MS{2137}, $z=\zcluster{m2137}$), and MACS~J0329-0211 (\M{0329}, $z=\zcluster{m0329}$) as blind test clusters, while the remaining clusters were organised into three different training sets based on different redshift ranges, as shown in Fig.~\ref{fig:exp2:layout}. Specifically: 
\begin{itemize}
    \item[-] \textit{Narrow}: clusters with redshift $0.332 \leq z \leq 0.412$ (514 \cm{s}, 555 \ncm{s})
    \item[-] \textit{Intermediate}: clusters with redshift $0.286 \leq z \leq 0.467$ (898 \cm{s}, 1157 \ncm{s})
    \item[-] \textit{Large}: clusters with redshift $0.174 \leq z \leq 0.606$ (1286 \cm{s}, 1725 \ncm{s})
\end{itemize}

The training set configurations were mostly organised to identify \cm{s} in \A{370}. This is the most significant test bench since it includes $535$ spectroscopic sources and is in the middle of \cm{} redshift range.
The other two clusters, \MS{2137} and \M{0329}, were chosen as additional test sets located at redshifts lying outside the \textit{narrow} and \textit{intermediate} ranges, while remaining well within the \textit{large} training set. 

\begin{figure*}[htbp]
   \centering
   \includegraphics[width=2\columnwidth]{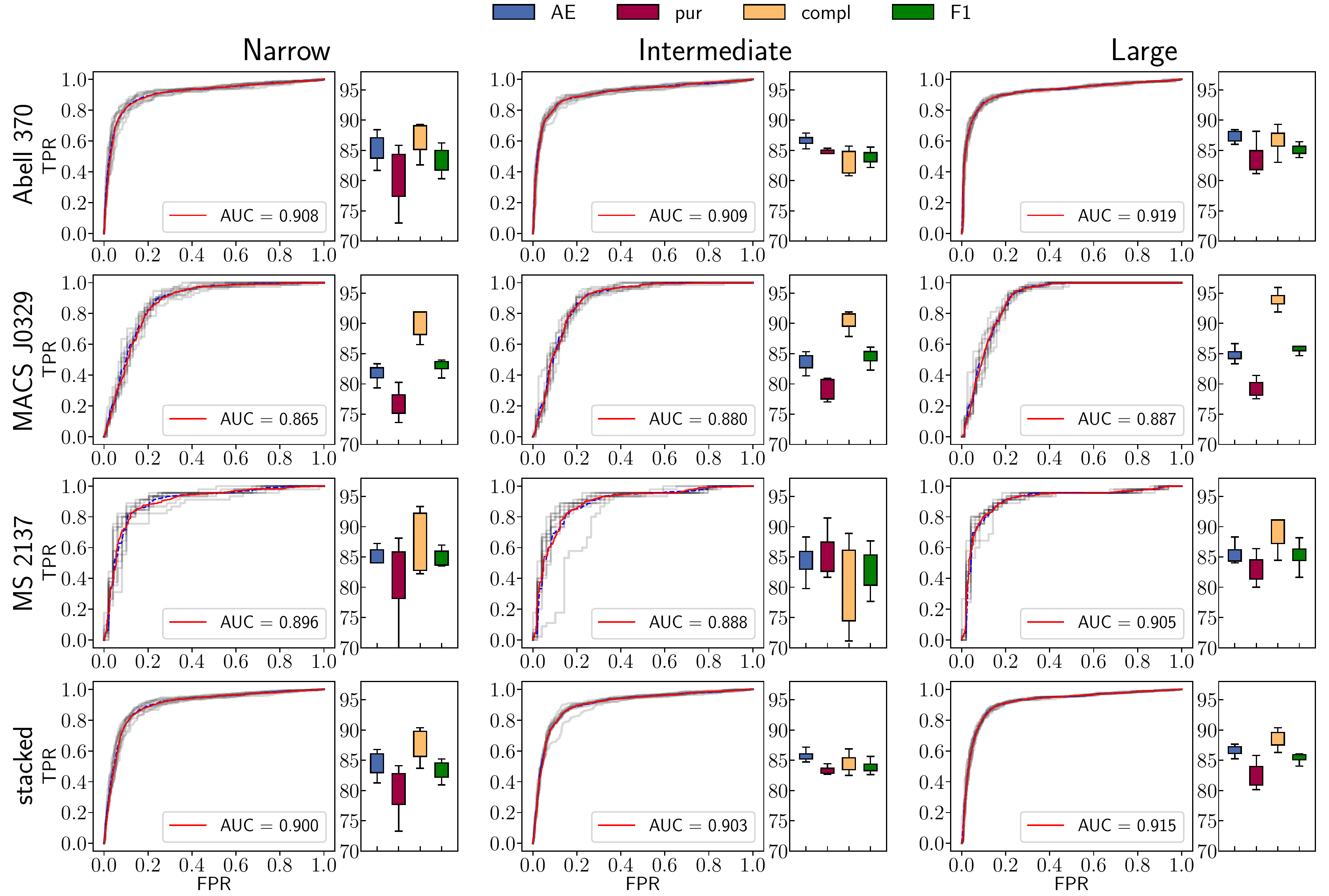}
   \caption{Summary of the \textit{EXP2} experiment. The statistical performances for the three  clusters (\A{370}, \M{0329} and \MS{2137}) are reported in each row, while results for the three training configurations (i.e. \textit{narrow}, \textit{intermediate} and \textit{large}) are organised by column. The global performances achieved by stacking together the three clusters are reported in the bottom row. For each test set, we display the ROC curves (grey lines refer to the performances achieved by any training fold, while the main trend is emphasised in red, together with its AUC score); the box plots represent the fluctuation of measured estimators related to the \cm{s}, together with the average efficiency measured for both classes. As in Fig.~\ref{fig:exp1:zbin}, such boxes delimit the $25$th and $75$th percentiles, while error bars enclose the maximum point variations.}\label{fig:exp2}
\end{figure*}

The results are shown Fig.~\ref{fig:exp2} and detailed in Table~\ref{tab:exp2}. They show that: (\textit{i}) the \textit{large} training set reached best results in most cases, with an average improvement between $1.1\%$ and $4.3\%$ with respect to the \textit{intermediate} case; (\textit{ii}) the \textit{narrow} training ensemble provided, in most cases, the worst results, showing a lower trade-off between purity and completeness, particularly evident (larger than $3\sigma$) for \A{370} and \M{0329}. This confirmed that the best performances were reached by extending the knowledge base, that is, when the \cm{} training sample covers the largest available redshift range.

\begin{table}[tbp]\caption{Statistical performances of the CNN model in EXP2. Best results are emphasised in bold. 
}
\label{tab:exp2:magcolsplit}\centering
\begin{tabular}{lcccc}
\hline
&\multicolumn{4}{c}{stacked}\\
\%& bright & faint & redder & bluer\\\hline
pur & $\mathbf{85.9\pm0.4}$ & $82.2\pm0.8$ & $\mathbf{91.0\pm0.5}$ & $79.4\pm0.9$\\ 
compl & $\mathbf{95.2\pm0.7}$ & $81.4\pm1.0$ & $\mathbf{95.2\pm0.6}$ & $75.7\pm1.0$\\ 
F1 & $\mathbf{90.3\pm0.4}$ & $81.7\pm0.8$ & $\mathbf{93.1\pm0.7}$ & $77.6\pm0.8$\\\hline
&\multicolumn{4}{c}{\A{370}}\\
\%& bright & faint & redder & bluer\\\hline
pur & $\mathbf{88.4\pm0.7}$ & $83.6\pm0.9$ & $\mathbf{90.5\pm0.7}$ & $79.8\pm1.0$ \\
compl & $\mathbf{96.8\pm0.7}$ & $80.8\pm1.2$ & $\mathbf{93.9\pm0.4}$ & $77.4\pm1.2$ \\
F1 & $\mathbf{92.4\pm0.7}$ & $82.1\pm0.9$ & $\mathbf{92.2\pm0.8}$ & $78.6\pm0.9$ \\\hline
&\multicolumn{4}{c}{\M{0329}}\\
\%& bright & faint & redder & bluer\\\hline
pur & $80.7\pm0.6$ & $\mathbf{81.1\pm1.7}$ & $\mathbf{88.3\pm0.9}$ & $74.4\pm1.2$ \\
compl & $\mathbf{98.0\pm1.0}$ & $85.1\pm0.6$ & $\mathbf{95.1\pm0.6}$ & $78.6\pm0.8$ \\
F1 & $\mathbf{89.3\pm0.5}$ & $83.0\pm1.7$ & $\mathbf{91.7\pm0.7}$ & $76.5\pm1.0$ \\\hline
&\multicolumn{4}{c}{\MS{2137}}\\
\%& bright & faint & redder & bluer\\\hline
pur & $\mathbf{90.8\pm1.0}$ & $76.7\pm1.5$ & $\mathbf{87.5\pm0.3}$ & $72.0\pm1.3$ \\
compl & $\mathbf{88.9\pm1.2}$ & $80.0\pm0.9$ & $\mathbf{90.6\pm0.6}$ & $76.2\pm1.0$ \\
F1 & $\mathbf{89.7\pm1.1}$ & $78.3\pm1.2$ & $\mathbf{89.0\pm0.4}$ & $74.1\pm1.1$ \\
\end{tabular}
\end{table}

We also analysed the CNN classification performances separately on bright and faint (\textit{EXP2a}) galaxies, as well as on red and blue galaxies (\textit{EXP2b}). The magnitude values adopted to split the \cm{} into equally sized samples are $F814$=22.0, 21.7, and 21.6 mag for \A{370}, \M{0329}, and \MS{2137}, respectively. For the analysis of the colour dependence, we used the $(F814 - F160)$ colour. However, since this colour depends on the $F814$ magnitude, we defined the difference between the observed colour and the colour-magnitude relation, that is, $(F814 - F160)_{\text{diff}} = (F814 - F160)_{\text{obs}} - [\text{colour-magnitude}(F814)]$. The colour-magnitude relation was fitted for each cluster with spectroscopic confirmed members, using a robust linear regression \citep{cappellari2013}, which is a technique that allows for a possible intrinsic data scatter and clips outliers, adopting the least trimmed squares technique \citep[][]{Rousseeuw2006}. By applying the correction for the colour-magnitude, we found that blue members can be defined as galaxies having $(F814 - F160)_{\text{diff}}<-0.160$, $-0.165$, $-0.157$ for \A{370}, \M{0329}, and \MS{2137}, respectively. Both experiments (a and b) were performed using the \textit{large} redshift configuration.

The results of the \cm{} identification are shown in Table~\ref{tab:exp2:magcolsplit}. In \textit{EXP2a}, all the statistical estimators indicated a very good performance of the method, although with a slightly lower efficiency in identifying faint objects. In fact, brighter members were detected with higher completeness ($90\%-98\%$) and purity ($81\%-91\%$), with a significant F1 score improvement ($89\%-92\%$), when compared to fainter members (completeness: $80\%-85\%$; purity: $77\%-85\%$;  F1 score: $78\%-83\%$), obtaining remarkable results for \A{370}, in which purity and completeness of \cm{s} are $\sim88\%$ and $\sim97\%$, respectively. Nevertheless, fainter \cm{s} were identified with an acceptable F1 score ($\sim80\%$). 

The experiment, \textit{EXP2b}, also showed good performances of the method for both red and blue objects, although the colour dependence of the results was evident. In particular, red galaxies were classified with a mean F1 score of $\sim91\%$, decreasing down to $\sim77\%$ for blue objects. The results reflect the underlying similarity between blue members and background objects, which implies that they cannot be separated easily. This was confirmed by the analysis of false positives and false negatives discussed in Sect.~\ref{sec:discussion}.

\subsection{EXP3: Comparison with photometric approaches}\label{ss:exp3}
This section is dedicated to a comparison of the classification performance of cluster members using the image-based DL method described above along with two different techniques based on photometric catalogues. The first is a random forest classifier (developed by our team) and the second one is a photometry-based Bayesian model described in \citealt{grillo2015} and in Appendix~\ref{app:benchmethods}.

\begin{figure}[htbp]
   \centering
   \includegraphics[width=0.98\columnwidth]{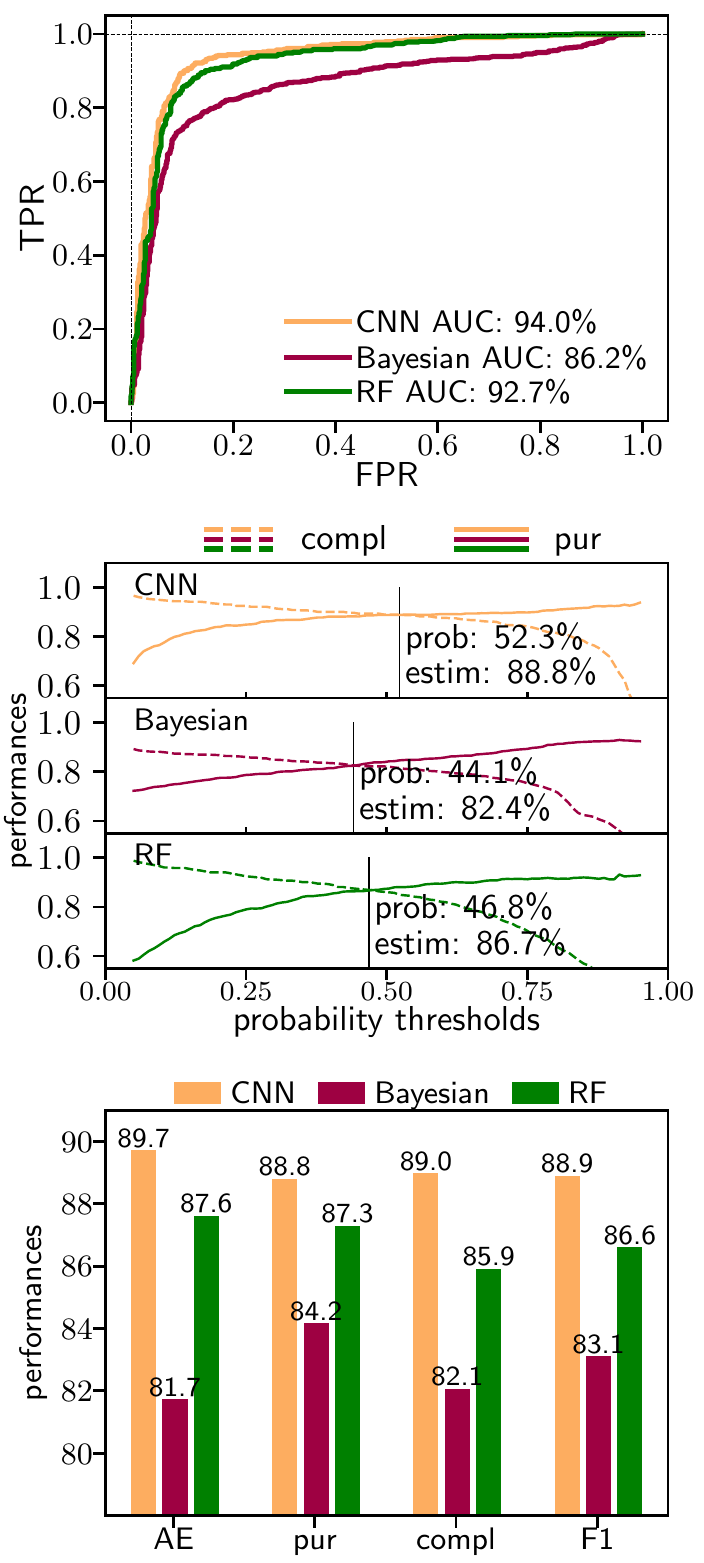}
   \caption{Comparison among the image-based CNN and two photometric catalogue-based approaches, namely, a random forest and Bayesian method (\textit{EXP3}), by combining results from the four clusters (\R{2248}, \M{0416}, \M{1206}, \M{1149}). Upper panel shows the ROC curves for the three methods with measured Area Under the Curve (AUC). The middle panel reports the trends of purity and completeness as a function of the probability thresholds used to obtain the ROC curves. In the three diagrams, we mark the intersection between such curves, i.e. the probability for which completeness and purity are equal. Bottom panel shows the differences between the three methods based on the statistical estimators described in Sect.~\ref{ss:estimators}.}\label{fig:exp3:roc}
\end{figure}

In this experiment, our CNN was trained with the \textit{mixed*} filter set (see Sect.~\ref{sec:data}). We focused on the results obtained by these three methods on \R{2248}, \M{0416}, \M{1206}, and \M{1149}. The statistical estimators are shown in detail in Table~\ref{tab:exp3} and in Fig.~\ref{fig:exp3:roc:clusters} as ROC curves, while in Fig.~\ref{fig:exp3:roc}, the performances are summarised by combining the results from the four clusters based on their ROC curves (top), the trade-off between purity and completeness (middle), and the usual statistical estimators (bottom). The photometric techniques show an average efficiency around $86-89\%$, with some values $\gtrsim 96\%$ for the Bayesian approach, although the F1 scores always remain between $83\%$ and $88\%$. The CNN confirmed its ability to detect \cm{s} with an F1 score between $87\%$ and $91\%$. The upper panel in Fig.~\ref{fig:exp3:roc} shows that globally CNN reaches an AUC of $\sim94\%$, which is $\sim8\%$ higher than the Bayesian method, while exhibiting the sharpest rise and the highest plateau. This means that for the CNN method there is a larger probability range in which the performances remain stable, while for the other methods a fine-tuning of the probability value is needed to balance purity and completeness. Furthermore, CNN reached the best trade-off between purity and completeness with a cross-over at $\sim89\%$. 
A summary of the results is shown in the bottom panel of Fig.~\ref{fig:exp3:roc}, where the differences among the CNN and the two photometric methods are measured using the four statistical estimators. The CNN performances were overall near $90\%$ and remained consistently higher than those of photometric-based methods. 
Finally, we analysed the common predictions among the three methods, both in terms of correctly classified and misclassified sources, separately for \cm{s} and \ncm{s}. Such results are graphically represented in Fig.~\ref{fig:exp3}. All three methods share $\sim76\%$ of their commonalities (i.e. summing of correct and incorrect predictions), of which, $\sim97\%$ (i.e. $74.6\%$ with respect to the whole set of common sources) were correctly classified. Common true positives and true negatives (i.e. \cm{s} and \ncm{s} that have been correctly classified) were  $\sim75\%$. The CNN and Bayesian method shared the largest fraction of predictions $\sim90\%$ (of which $\sim93\%$ were correct) with respect to the joint classification of CNN and RF ($\sim82\%$); this implied that RF had a significant fraction of uncommon predictions ($\sim 14\%$). 

Concerning the misclassified objects, the methods shared $\sim2\%$ of incorrect predictions, of which: $\sim1\%$ of \cm{s} were common false negatives (FNs, i.e. \cm{s} sources wrongly predicted as \ncm{s}), while $2.5\%$ were common false positives (FPs, i.e. \ncm{s} sources wrongly predicted as \cm{s}). The CNN exhibited the least fraction of misclassifications (about $10\%$). 
The CNN showed a percentage of FNs larger than Bayesian ($10\%$ versus $7\%$), which, in turn, had a wider FP rate ($11\%$ versus $17\%$). Therefore, although CNN and Bayesian methods shared a significant fraction of incorrect predictions ($85\%$ of common misclassifications, suggesting the existence of a fraction of sources for which the membership is particularly complex for both of them), these two models exhibit a different behaviour: the CNN tended to produce more pure than complete \cm{s} samples, whereas the Bayesian method showed the opposite, which is in agreement with what is reported in Table~\ref{tab:exp3}.

\section{Photometric selection of \cm{s}}\label{sec:run}

The experiments described in the previous sections are mostly focused on the classification efficiency and limits of the image-based CNN approach and evaluating its dependence from observational parameters such as redshift, number of \cm, photometric band compositions, magnitude, and colour. In this section, we are mainly interested in evaluating the degree of generalisation capability of the trained CNN in classifying new sources as cluster members, a step process that is commonly referred to as \textit{run} in the ML context.

In particular, we applied the CNN model to the photometrically selected \cm{s} in \R{2248}, \M{0416}, \M{1206}, and \M{1149}. The training set was constructed by combining all clusters with the \textit{mixed*} band configuration, using the k-fold approach (see Sect.~\ref{sec:method}).

Similarly to what was done to build the knowledge base (see Sect.~\ref{sec:data}), for the \textit{run} set we used squared cut-outs $\sim4\arcsec$ across, centered on the source positions as extracted by SExtractor \citep{bertin:1996}. Thus, the \textit{run} set was composed by $5269$ unknown sources, of which $1286$, $1029$, $1246$, and $1708$ were in the FoV of \R{2248}, \M{0416}, \M{1206}, and \M{1149}, respectively.

\begin{figure*}[htbp]
   \centering
   \includegraphics[width=2\columnwidth]{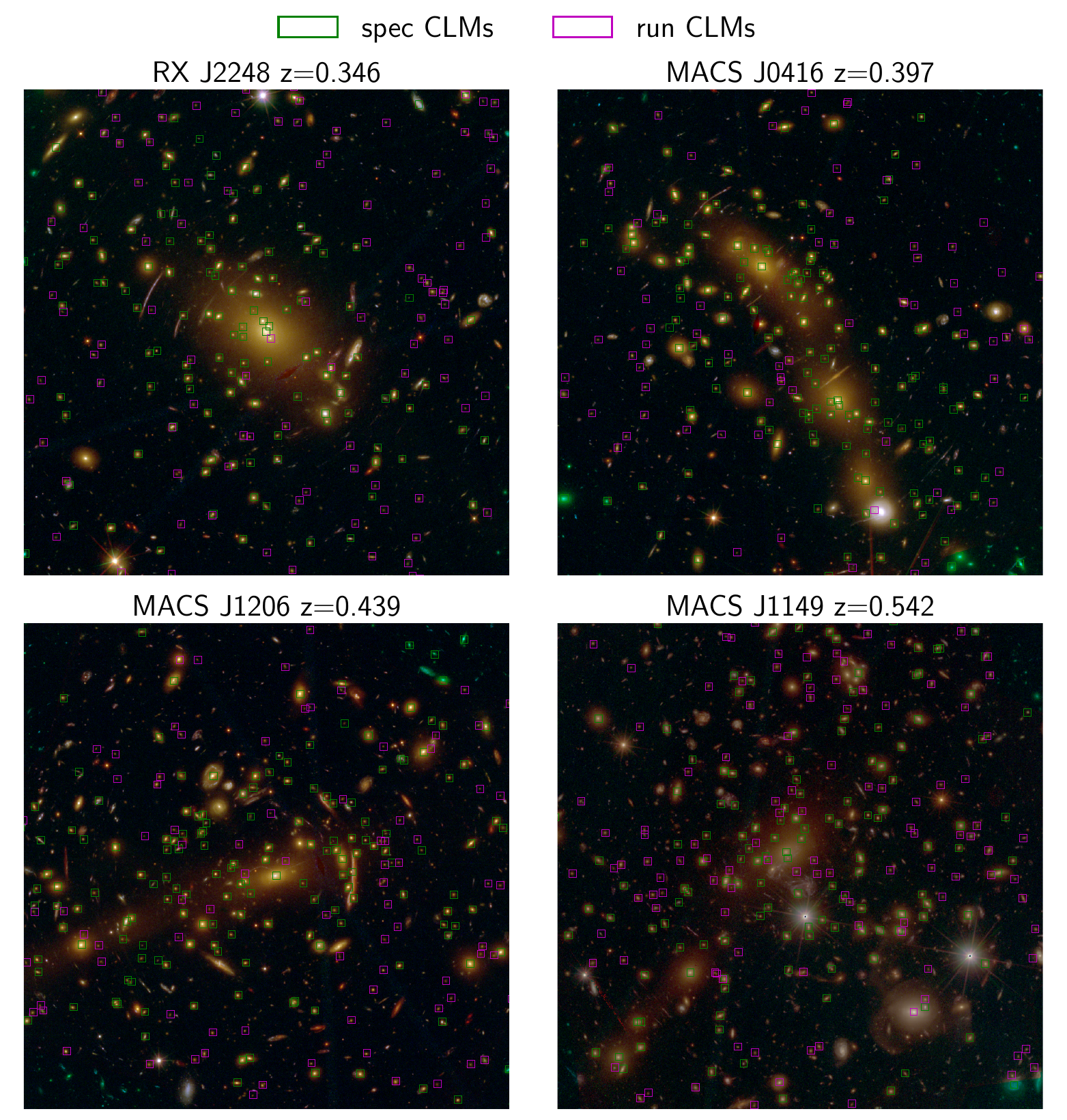}
   \caption{CNN member selection (marked with open magenta squares) obtained with the \textit{run} set, together with the spectroscopic \cm{s} (marked with open green squares), in the core of the four clusters \R{2248} ($z=\zcluster{r2248}$), \M{0416} ($z=\zcluster{m0416}$), \M{1206} ($z=\zcluster{m1206}$) and \M{1149} ($z=\zcluster{m1149}$). All images are $130$ arcsec across.}\label{fig:exp4:clusters}
\end{figure*}

\begin{figure*}[htbp]
   \centering
   \includegraphics[width=2\columnwidth]{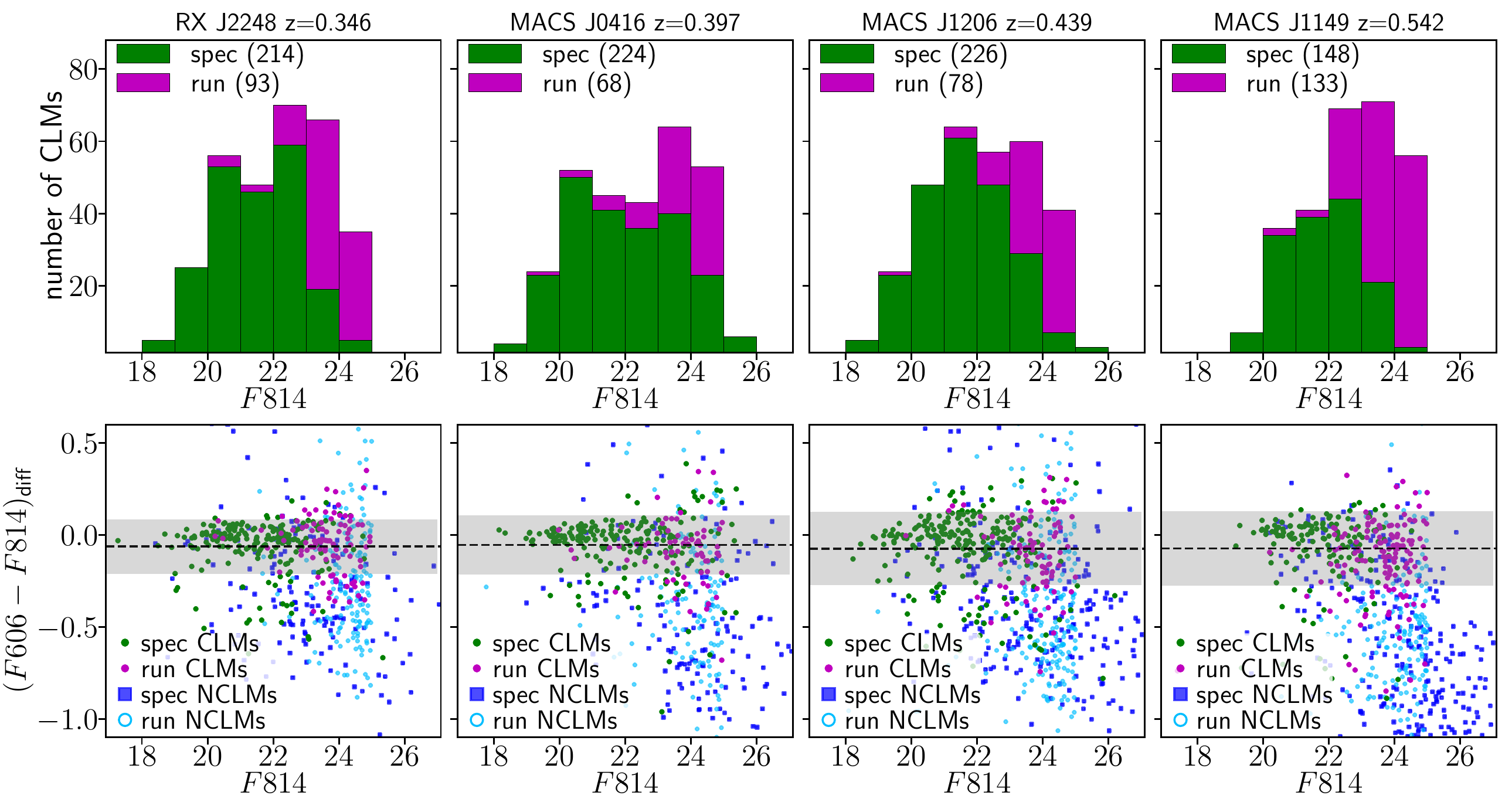}
   \caption{CNN membership prediction (\textit{run}) together with spectroscopic sources, represented as (i) \cm{s} distribution of $F814$ magnitudes (first row), (ii) differential colour - magnitude sequence for both \cm{s} and \ncm{s}. Spectroscopic \cm{s} are shown in green, candidate members in purple, spectroscopic \ncm{s} with blue squares and candidate \ncm{s} with open cyan circle. We only plot CNN cluster members with $F814 \le 25$mag. The grey region within the CM diagrams limits the area corresponding to $\pm1\sigma$ from the median (dashed horizontal line) of $(F606-F814)_{\text{diff}}$.}\label{fig:exp4}
\end{figure*}

The CNN identified a total of $372$ members with $F814 \le 25$ mag, which is approximately the magnitude limit of the spectroscopic members (only $\sim 3\%$ of spectroscopic members has $F814> 25$), with $\sim46\%$ of candidate \cm{s} having membership probabilities larger than $90\%$. The spatial distribution of both spectroscopic and predicted \cm{s} are shown in Fig.~\ref{fig:exp4:clusters}, while the magnitude ($F814$) distribution and the colour-magnitude relations ($F606-F814$ versus $F814$) for both spectroscopic and predicted members are shown in Fig.~\ref{fig:exp4}. The magnitude distributions indicate that the CNN was able to complete the spectroscopic \cm{s} sample down to $F814=25$. This was also confirmed by the analysis of the colour-magnitude diagrams, which show that the photometrically identified \cm{s} complete the spectroscopic red-sequence at $F814<25$, emphasising the CNN capability to disentangle \cm{s} from background objects. We counted also the number of recognised \cm{s} within, respectively, $1$, $2$, and $3\sigma$ from the median of differential colour $(F606-F814)_{\text{diff}}$.

\section{Discussion}\label{sec:discussion}

One particular aspect that is often addressed when using ML methods is the impact on the classification performances carried by the amount of data available, both in terms of the number of features (photometric bands) and amount of training objects. The \textit{EXP1} (see Sect.~\ref{ss:exp1}) enabled an analysis of the trade-off between the information carried by the imaging bands and the number of samples in the dataset. As reported in Table~\ref{tab:exp1:stack} and Table~\ref{tab:exp1:clusters}, there was a small improvement of efficiency ($\sim3\%$) by increasing the size of the training sample by $34\%$, when comparing the two \textit{mixed} and \textit{mixed*} configurations. However, these two data samples included both optical and infrared information. To better understand this important aspect, we performed a comparison between data samples with and without the infrared bands. Such analysis was carried out by directly comparing the two \textit{ACS} and \textit{ALL} configurations, although the sample size of the second one was $\sim30\%$ smaller. The results, shown in Table~\ref{tab:exp1:stack} and Table~\ref{tab:exp1:clusters}, suggested that the addition of infrared imaging adequately compensated the smaller size of the training set. 

We also investigated the dependence of member classification performance on the magnitudes and colours. Here, the \textit{EXP2a} showed very good performances of the method for both bright and faint sources, although with a slightly lower efficiency in identifying fainter objects. On the other hand, \textit{EXP2b} showed a mean efficiency of $\sim$ 91\% in classifying red galaxies, which was reduced to $\sim$ 77\% for blue objects (see Table~\ref{tab:exp2:magcolsplit}). 

\begin{figure*}[htbp]
   \centering
   \includegraphics[width=2\columnwidth]{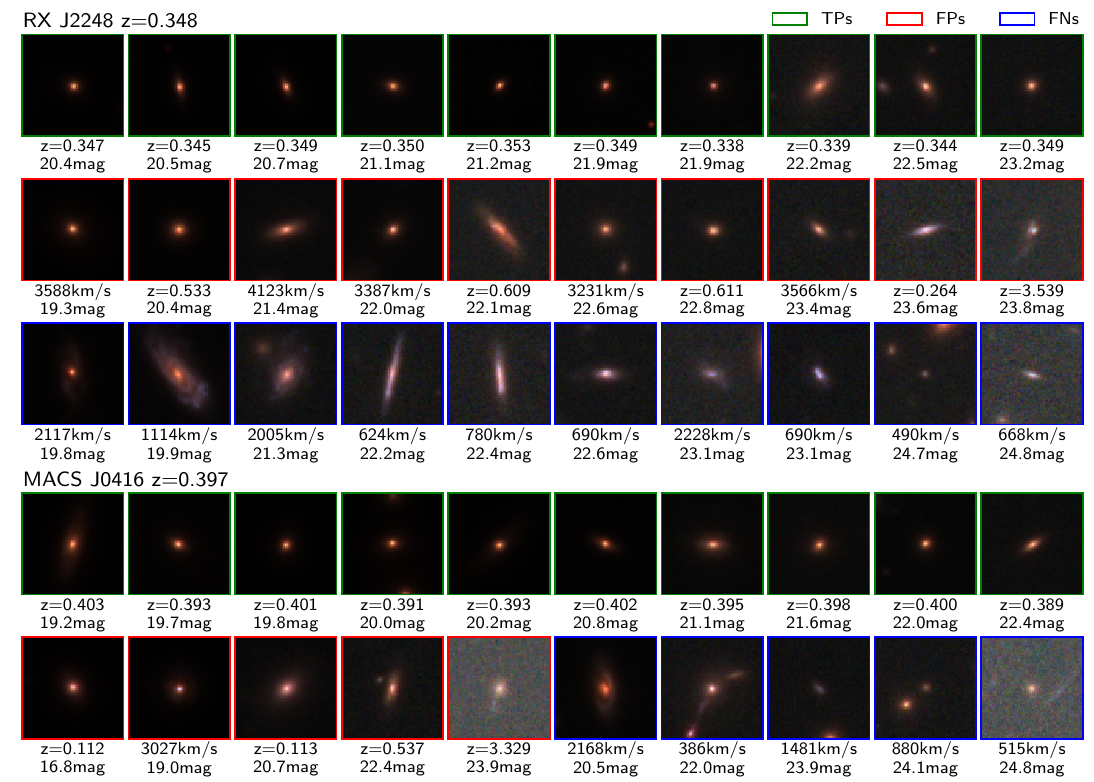}
   \caption{Ensemble of object cut-outs with a size of $64$ pixels ($\sim4\arcsec$), corresponding to some specific CNN predictions in the clusters \R{2248} (first three rows) and \M{0416} (last two rows). The True Positives (TPs), i.e. the \cm{s} correctly identified, are shown on first and fourth row with green boxes, while False Positive and False Negative (FPs and FNs) are shown on the second, third and fifth row, framed by red and blue boxes, respectively. The images were obtained by combining five HST bands: $F435$, $F606$, $F814$, $F105$, $F140$. The figure shows sources in the $F814$ band with a magnitude $F814\leq25$mag. TPs are shown together with their spectroscopic redshift, while FNs together with their cluster rest-frame velocity separation. For convenience, in the case of FPs, their cluster velocity separations are quoted when within $\pm9000$km/s, otherwise their redshift is shown.}\label{fig:tp_fp_fn_1}
\end{figure*}

\begin{figure*}[htbp]
   \centering
   \includegraphics[width=2\columnwidth]{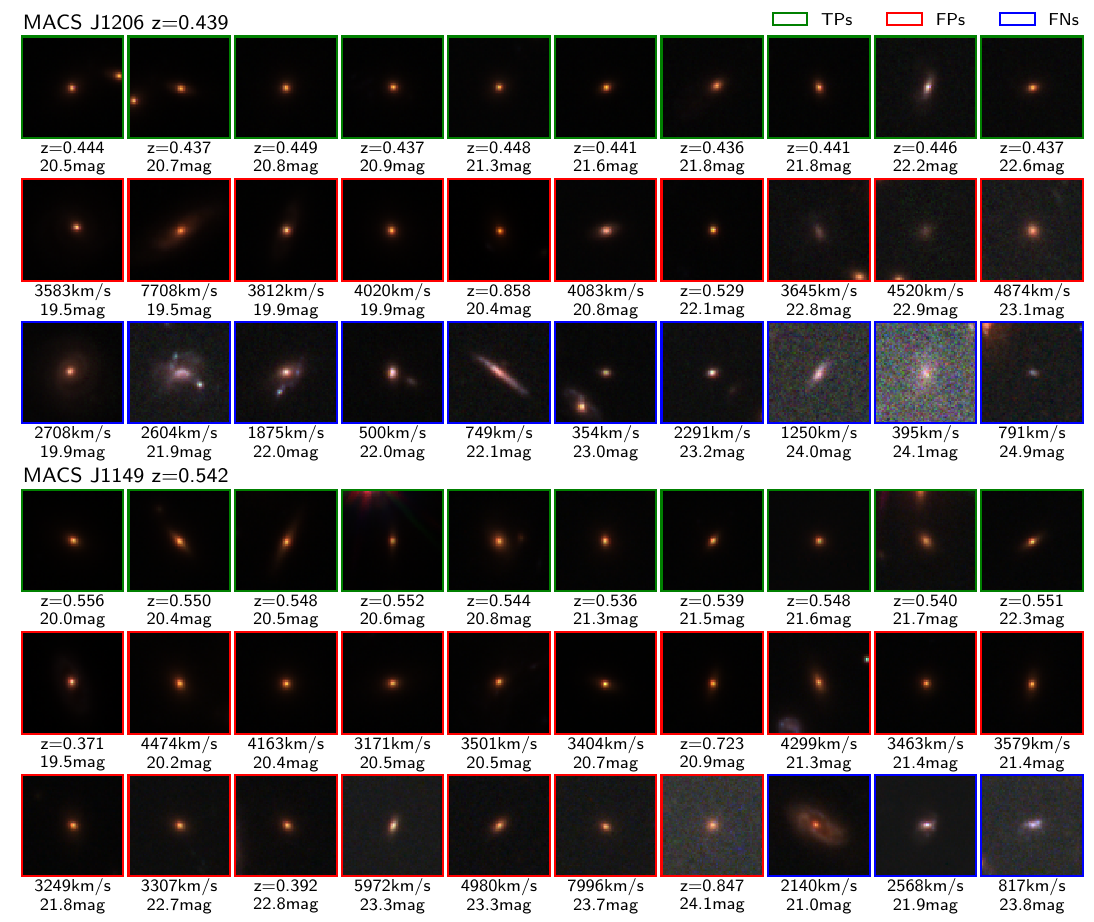}
   \caption{Same of Fig.~\ref{fig:tp_fp_fn_1} for the cluster \M{1206} (first three rows) and \M{1149} (last three rows).}\label{fig:tp_fp_fn_2}
\end{figure*}

\begin{figure}[htbp]
   \centering
   \includegraphics[width=\columnwidth]{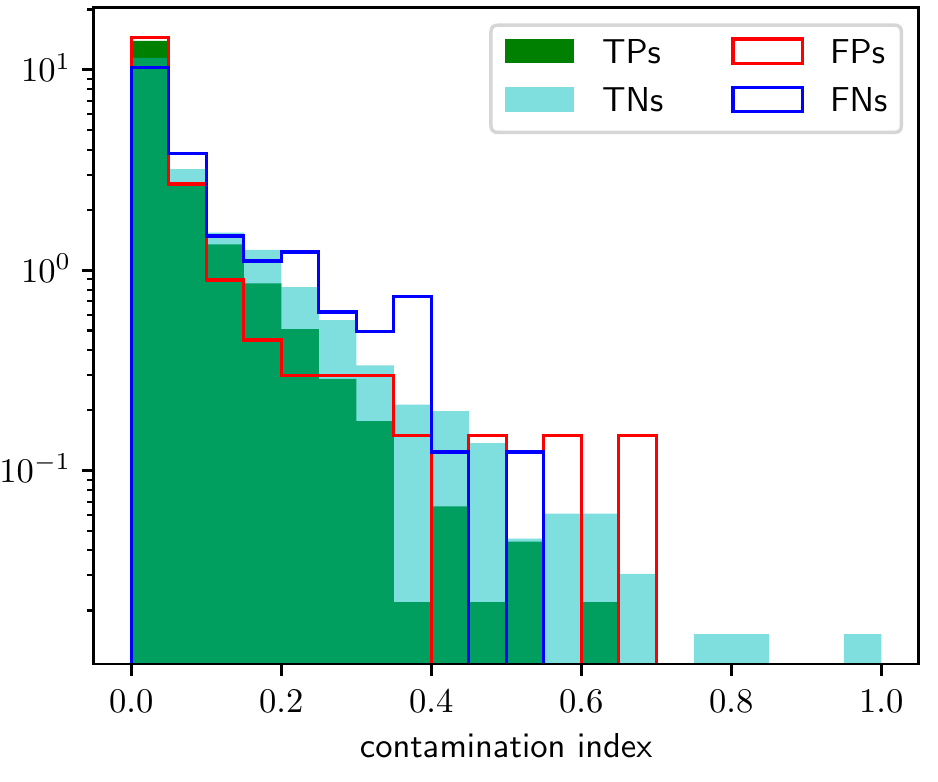}
   \caption{Logarithmic distribution of the \textit{contamination index} for true positives (TPs, green), true negatives (cyan), false positives (red), and false negatives (blue). The distribution includes all available clusters. }\label{fig:cont_index}
\end{figure}

\begin{figure*}[htbp]
   \centering
   \begin{subfigure}[t]{0.856\columnwidth}
        \includegraphics[width=\linewidth]{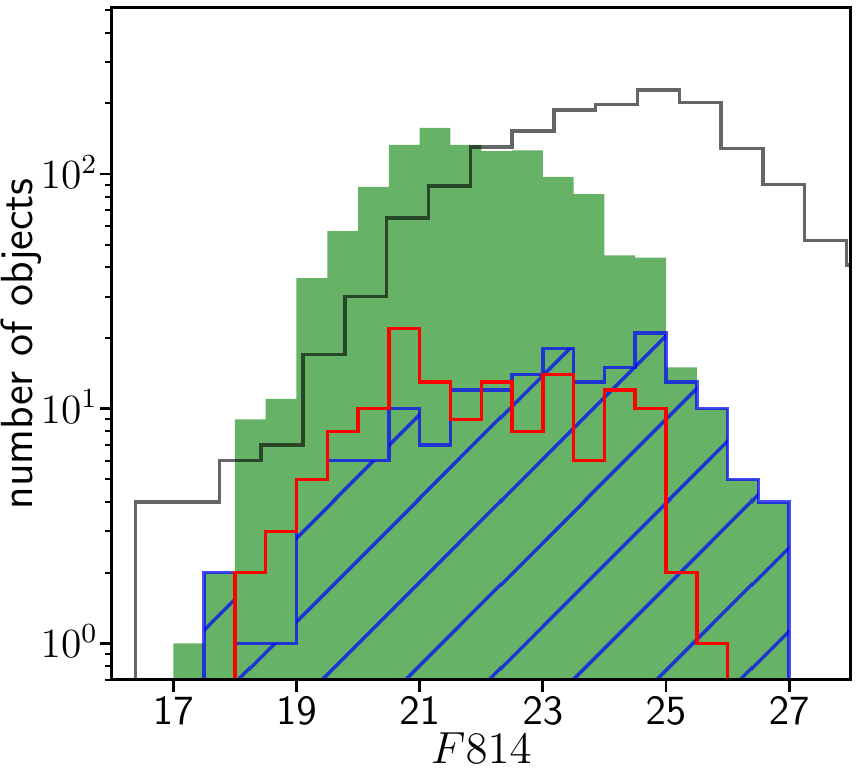}\caption{}\label{fig:fp_fn_stacked:mag}
   \end{subfigure}
   \begin{subfigure}[t]{1.144\columnwidth}
        \includegraphics[width=\linewidth]{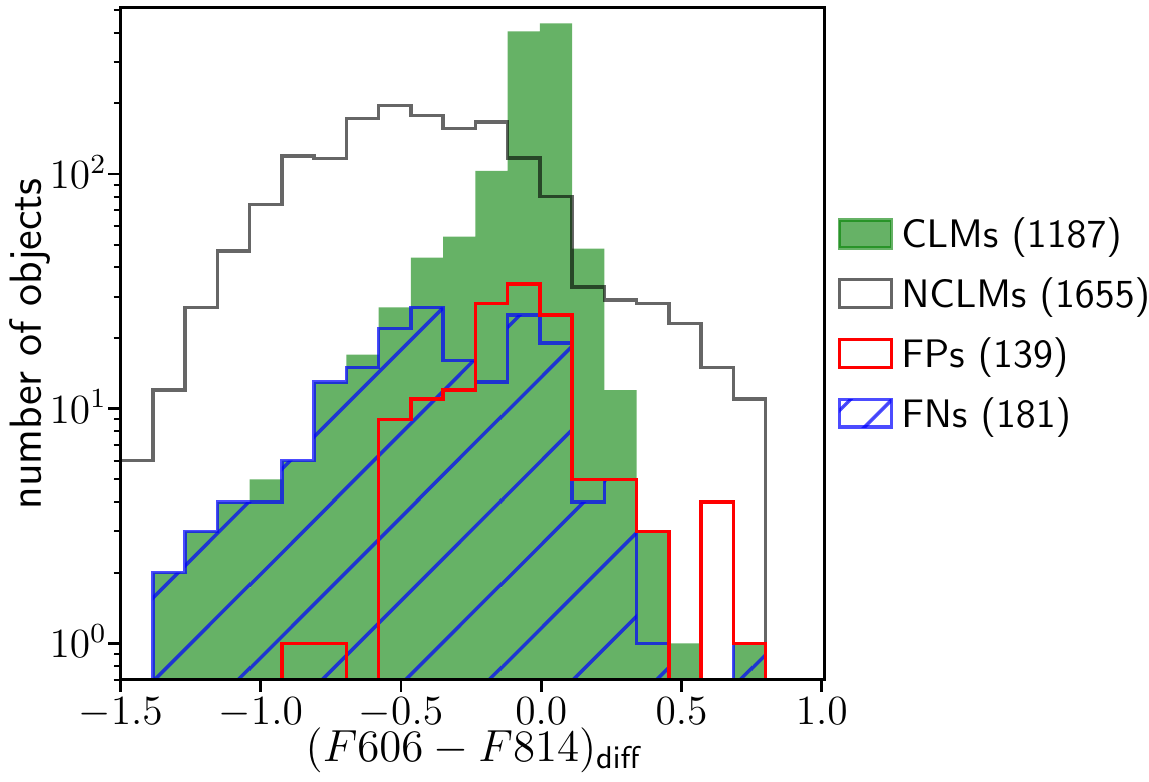}\caption{}\label{fig:fp_fn_stacked:col}
   \end{subfigure}
   \caption{Magnitude (left panel) and colour (right panel) logarithmic distributions of FPs (red) and FNs (blue), overlapped to the \cm{} (green) and \ncm{} distributions, for the fifteen clusters (\textit{stacked}) included in our study. The number of objects for each plotted distribution is quoted in brackets in the legend. The differential colour $(F606-F814)_{\text{diff}}$ is obtained by applying the correction for the mean colour-magnitude relation for each cluster. Table~\ref{tab:col_fp_fn} outlines such results.}\label{fig:fp_fn_stacked}
\end{figure*}

\begin{figure*}[htbp]
   \centering
   \includegraphics[width=0.69\textwidth]{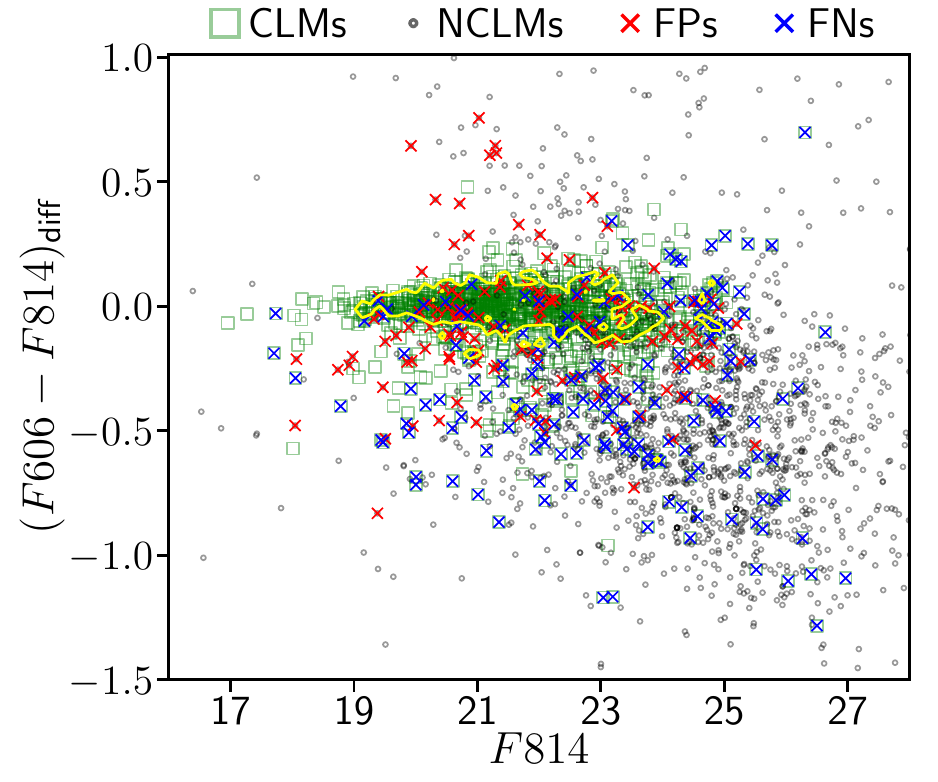}
   \caption{Colour-magnitude relation for the \cm{s} (green squares), with the overlapped distributions of FPs (red crosses), FNs (blue crosses) and \ncm{s} (grey circles), for the sample of fifteen clusters (\textit{stacked}). The yellow contour delimits the red-sequence at $1\sigma$ confidence level. Colours reported on the y-axis are corrected for the mean red-sequence of each cluster (see Sect.~\ref{sec:exps}).}
   \label{fig:fn_fp_stacked:colmag}
\end{figure*}

To further investigate the robustness in the identification of cluster members (i.e. the \textit{positive} class), from the classification confusion matrices, we defined true positives (TPs) the \cm{s} correctly classified, false positives (FPs) \ncm{s} classified as \cm{s}, false negatives (FNs) the \cm{s} classified as \ncm{s}, and, finally, true negatives (TNs) as \ncm{s} correctly classified. A short sample of TPs, FPs and FNs in \R{2248} and \M{0416}, and in \M{1206} and \M{1149} are shown in Fig.~\ref{fig:tp_fp_fn_1} and  Fig.~\ref{fig:tp_fp_fn_2}, respectively. We explored the model predictions, by inspecting the TPs and the distribution of FPs and FNs as function of their magnitude and colour.  

A critical aspect of the classification of members within the central cluster region is the impact of crowding. Therefore, we specifically focused on the DL ability to predict cluster membership in such circumstances (see a few examples of cut-outs in Figs.~\ref{fig:tp_fp_fn_1} and~\ref{fig:tp_fp_fn_2}).

We introduced a contamination index ($CI$) for each cut-out, defined as: $CI=\sum_{i=1}^{N_c}1/(d_i\cdot F814_i)$, where $N_c$ is the number of contaminants in the cut-outs, $d_i$ is the distance in arcsec between the central source and $i-$th contaminant, while $F814_i$ is the magnitude of the contaminating source. The indices for cut-outs without contaminants were set to zero. Then, we normalised this index in the $[0,\,1]$ interval. Fig.~\ref{fig:cont_index} shows that the four contamination index distributions of, respectively, TPs, TNs, FPs and FNs mostly overlapped and followed the same trend. 
In fact, the $48\%$ of FNs and $28\%$ of FPs had a non-zero contamination index, as well as the $31\%$ and $43\%$ of TNs and TPs. The lack of a correlation between the contamination index and incorrect prediction rates (FPs and FNs) suggests that the source crowding did not significantly affect the CNN classification efficiency. 

By analysing the FP and FN rows in Figs.~\ref{fig:tp_fp_fn_1} and~\ref{fig:tp_fp_fn_2}, we can see an interesting dichotomy: FPs appear as red galaxies, while FNs as blue; in addition, the FPs have $F814<24$, whereas FNs are found also down to $F814\sim25$. In order to quantify such behaviours, we analysed the distribution of FPs and FNs in terms of: \textit{(i)} the $F814$ magnitude for both FPs and FNs (Fig.~\ref{fig:fp_fn_stacked:mag}); \textit{(ii)} the correlation between the CNN incorrect predictions and differential colours $(F606-F814)_{\text{diff}}$ (Fig.~\ref{fig:fp_fn_stacked:col}). These results are summarised in Table~\ref{tab:col_fp_fn}.

\begin{table}[htbp]\caption{Summary of FP and FN distributions.} \label{tab:col_fp_fn}\center
\begin{tabular}{lccc}\hline
                                  & \bf{CLMs} & FPs  & FPs/NCLMs\\\hline 
Total Number                      & 1187   & 139    & 0.084\\\hline
$F814<$25.0                       & 96.6\% & 97.8\% & 0.131\\ \hline
$F814\ge$25.0                     &  3.4\% &  2.2\% & 0.005\\ \hline
$(F606-F814)_{\text{diff}}<-$0.5  & 5.8\%  &  4.3\% & 0.008\\
$(F606-F814)_{\text{diff}}<$0.0   & 60.8\% & 70.5\% & 0.070\\
$(F606-F814)_{\text{diff}}\ge$0.0 & 39.2\% & 29.5\% & 0.161\\\hline
\hline
                                  & \bf{NCLMs} & FNs & FNs/CLMs \\\hline
Total Number                      & 1655   & 181    & 0.152\\\hline
$F814<$25.0                       & 62.7\% & 79.0\% & 0.125\\ \hline
$F814\ge$25.0                     & 37.3\% & 21.0\% & 0.950\\ \hline
$(F606-F814)_{\text{diff}}<-$0.5  & 43.2\% & 35.4\% & 0.928\\
$(F606-F814)_{\text{diff}}<$0.0   & 84.6\% & 83.4\% & 0.209\\
$(F606-F814)_{\text{diff}}\ge$0.0 & 15.4\% & 16.6\% & 0.065\\\hline
\end{tabular}
\tablefoot{Fractions of \cm{s} (Col.~2), False Positives (FPs) (Col.~3) and the ratio of FPs to NCLMs (Col.~4) as a function of magnitude (\textit{second and third row}) and colours (\textit{fourth to sixth row)}. The total number of spectroscopic \cm{s} and FPs are quoted in the first row. Fractions as a function of colours are computed only for sources whose $F814$ and $F606W$ magnitudes are available ($\sim 84\%$ of the whole dataset). Similar fractions for \ncm{s}, FNs (False Negatives) and FNs/CLMs are quoted in the bottom half of the table.}
\end{table}

 Fig.~\ref{fig:fp_fn_stacked:mag} and Col.~4 in Table~\ref{tab:col_fp_fn} showed that almost all \cm{s} fainter than $F814W=25$ (representing a small fraction with respect to the total, see Col.~2 in Table~\ref{tab:col_fp_fn}) were FNs. This was not due to any failure on the part of the model, but, rather, to the poor sampling of such objects within the parameter space available to train the model. This was also confirmed when comparing the percentage of FPs and FNs with respect to the percentage of \cm{s} and \ncm{s} in Table~\ref{tab:col_fp_fn} as a function of magnitude. In fact, Table~\ref{tab:col_fp_fn} showed that the model tried to reproduce the distribution in terms of fractions of \cm{s} for FPs, and in terms of the fraction of \ncm{s} for FNs.

Finally, we analysed the correlations between the CNN incorrect predictions and colours. These distributions are shown in Fig.~\ref{fig:fp_fn_stacked:col} using the differential colour $(F606-F814)_{\text{diff}}$, while, in Table~\ref{tab:col_fp_fn}, the misclassification percentages are summarised. Also in this case, the distributions of FPs and FNs as a function of colours, are mimicking, respectively, the distributions of \cm{s} and \ncm{s} in Table~\ref{tab:col_fp_fn}. 

Very blue sources ($(F606-F814)_{\text{diff}}<-$0.5) populated only 5$.8\%$ of \cm{s} and represented the $\sim35.4\%$ of incorrect predictions, which is very similar to the fraction of very blue sources in the population of \ncm{s} (i.e. $43.2\%$). Conversely, redder sources were typically correctly classified, showing a FN rate of $16.6\%$. Moreover, from the fraction of FN/CLMs, we observed that almost all the blue cluster members were wrongly classified as \ncm{s} (see Col.~4 in Table~\ref{tab:col_fp_fn} and Fig.~\ref{fig:fp_fn_stacked:col}).

Regarding FPs, there was not a real classification problem with faint and very blue objects, whose rates in terms of \cm{s} were, respectively, $3.4\%$ and $5.8\%$, corresponding to $2.2\%$ and $4.3\%$ of incorrect predictions, respectively. From Table~\ref{tab:col_fp_fn}, it was also evident that within red misclassifications, FPs were more frequent than FNs ($29.5\%$ versus $16.6\%$), reproducing the distributions of \cm{s} ($39.2\%$) and \ncm{s} ($15.4\%$), respectively.

Figure~\ref{fig:fn_fp_stacked:colmag} shows the colour-magnitude relation of \cm{s} (green squares), overlapping the FP (red cross), FN (blue cross) and \ncm (grey circle) distributions. It emphasises the \cm{s} undersampling of the blue and faint region, together with the large concentration of FNs among bluer and fainter sources (see blue crosses). Among all the FNs, $\sim35\%$ are very blue ($(F606-F814)_{\text{diff}}<-0.5$), $\sim40\%$ of these had $F814>25$mag, suggesting that in the bluer region the FNs follows the \ncm{} distribution, while among FPs, $\sim64\%$ of them are red ($(F606-F814)_{\text{diff}}>-0.1$), but only $\sim1\%$ of these have magnitude fainter than $F814>25$mag. On the other hand, $\sim35\%$ of all FPs were within the yellow contours, which refer to the $1\sigma$ colour-magnitude relation, indicating that they were on the red sequence.

\begin{table}[htbp]\caption{Comparison among CNN performances considering the whole sample (Col.~2) and by removing sources with $F814\ge25$ and $(F606-F814)_{\text{diff}}<-$0.5 (Col.~3).}\label{tab:perf_final}\centering
\begin{tabular}{lcc}
\hline\hline
& Complete sample & $F814<$25.0 \& \\
&                 & $(F606-F814)_{\text{diff}}\ge-$0.5 \\\hline
true CLMs &	1187     & 1100 \\
pred CLMs &	1145     & 1130 \\
TPs	     &  1006     &  999 \\
FPs	     &   139     &  131 \\
FNs      &	 181     &  101 \\
pur      &	 87.9\%  & 88.4\%\\
compl	 &   84.8\%  & 90.8\%\\
F1	     &   86.3\%  & 89.6\%\\
\hline
\hline
\end{tabular}
\end{table}

In order to understand the impact of this misclassification of faint and very blue sources, we report, in Tab.~\ref{tab:perf_final}, the statistical estimators for the stacked sample and, individually for \R{2248}, \M{0416}, \M{1206}, \M{1149}), considering either the whole sample or by removing sources with $F814>25$ and very blue objects, that is, with $(F606-F814)_{\text{diff}}<-$0.5. 
By comparing these results, we observed a relevant increase of the completeness (for the stacked sample, it goes from 84.8\% to 90.8\%). This was mainly motivated by the sensible reduction of the FNs amount, which, by definition, had a higher impact on the completeness, rather than on other estimators. In fact, the purity and F1 score showed a smaller improvement, going, respectively, from 87.9\% to 88.4\% and from 86.3\% to 88.9\%. 

In summary, the FNs were mainly blue and faint. This was expected, given their under-representation in the dataset and their similarity with \ncm{s}. We note, in fact, that we were mapping a population of cluster members in the central and highest density region of clusters, dominated by a high fraction of bright and red members. Nevertheless, the simple exclusion of fainter sources with $F814>25$ and $(F606-F814)_{\text{diff}}<-0.5$ improved the CNN performance. Similar performances in terms of the distribution of false positives and negatives for sources with $F814>25$ and $(F606-F814)_{\text{diff}}<-0.5$ were obtained by the random forest classifier and the photometry-based Bayesian method. By comparing the behaviour of these three models on four clusters (R2248, M0416, M1206 and M1149), we found that the rate of blue FN is $28\%$ for the Bayesian method and $25\%$ for the random forest versus the $20\%$ for the CNN. The rate of faint FN is $1\%$ for the random forest and $6\%$ for the Bayesian method versus the $5\%$ of CNN. For what concerns FPs, the CNN, being the purest method, preserved the lowest contamination for both bluer and fainter members, with only four \ncm{s} classified as \cm{s}, compared with the $12$ and $24$ \ncm{s} for the Bayesian method and the random forest, respectively.

This comparison, while it confirms the good performances of the CNN, also shows that the three methods have comparable efficiencies in the faint and blue region of the parameter space, which is likely due to undersampling of members in this region of the knowledge base, as pointed out above. This is due to the fact that the population of galaxies in the densest central cluster regions is brighter and redder than that of the less dense and outer cluster regions (see \citealt{annunziatella2014, mercurio2016} for the specific study of M1206). Clearly, an improvement of the model's performances would require including member galaxies in the outer cluster regions and balancing the number of bluer and fainter members. In our case, even if the spectroscopic data cover more than two cluster virial radii, multi-band HST imaging with sufficient depth is only available in the central cluster regions.

\begin{figure*}[htbp]
   \centering
   \includegraphics[width=\columnwidth]{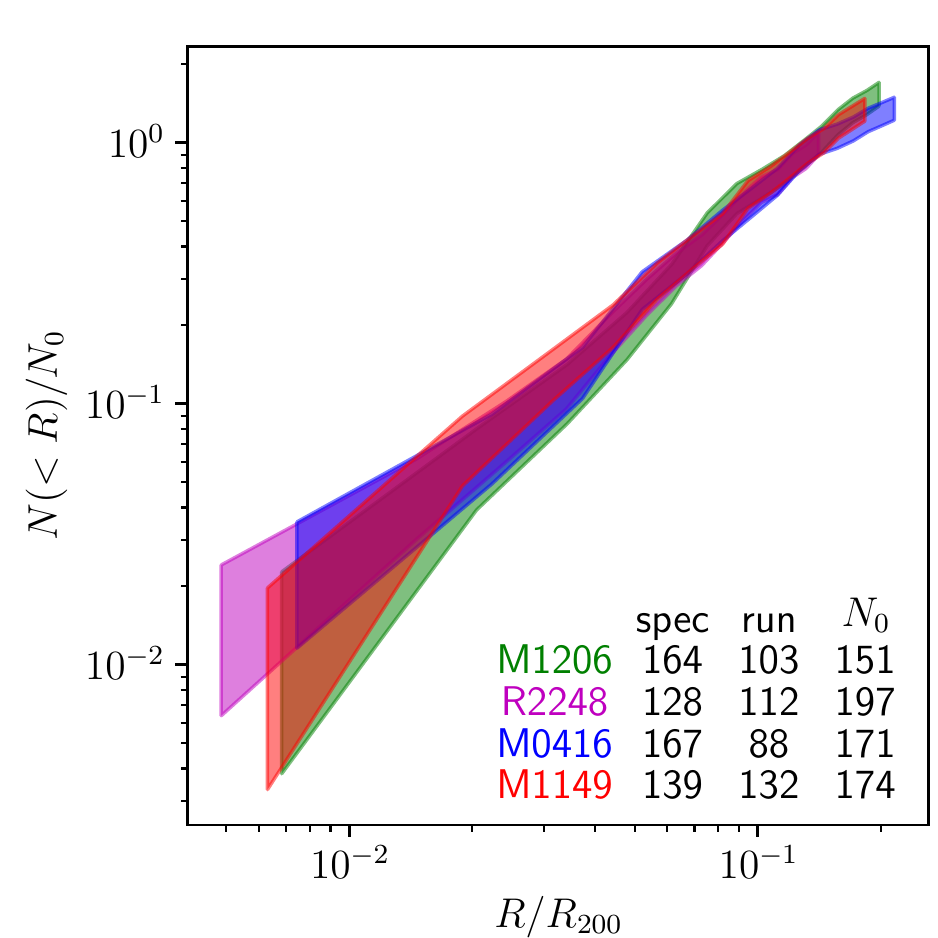}
   \includegraphics[width=\columnwidth]{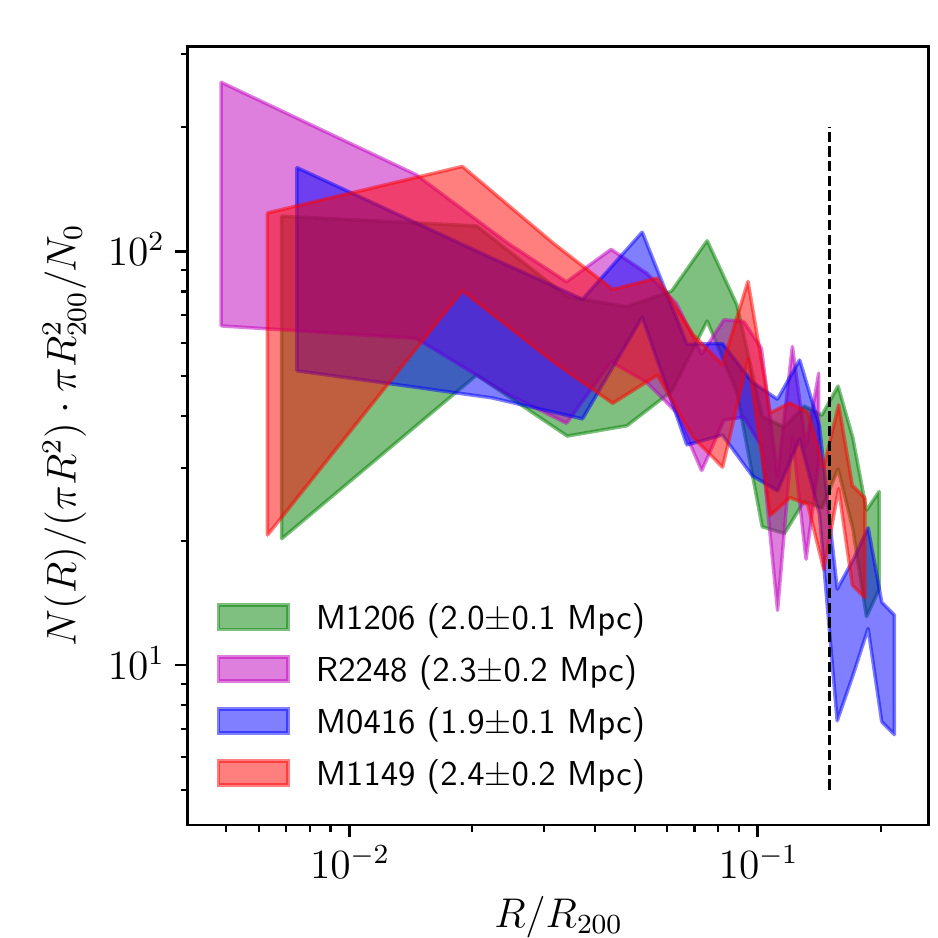}
   \caption{Cumulative (left) and differential (right) projected number of \cm{} for the four clusters (\R{2248}, \M{0416}, \M{1206}, and \M{1149}), including spectroscopic \cm{s} and candidate members identified by CNN (limited to $F814\leq25$ mag). The areas correspond to the $68\%$ confidence level regions. All profiles are normalised by the number $N_0$ of members with $R< 0.15\,R_{200}$ in all clusters. The number of spectroscopic, CNN-identified members ("run"), and $N_0$ values are quoted in the left panel. The adopted values of $R_{200}$ are quoted in the right panel, the computed values of $N_0$ are quoted in the left panel, together with the corresponding numbers of spectroscopic and "run" members. The dashed line in the right panel corresponds to $R=0.15R_{200}$.}\label{fig:density2d}
\end{figure*}

Finally, we used both spectroscopic members and candidate \cm{s} identified by CNN to estimate the cumulative projected number of cluster members and the differential number density profiles (Fig.~\ref{fig:density2d}). According to our previous analysis, we excluded candidate \cm{s} with $F814>25$ mag, where only $\sim\!3 \%$ of spectroscopic members were present. To properly compare profiles of clusters with different virial masses, we computed the values $R_{200}$ from of the values of $M_{200c}$ obtained by \citet{umetsu2018} with independent weak lensing measurements\footnote{We note again that $R_{200} = \left(\frac{2G}{H(z_{cl})^2} \frac{M_{200c}}{200}\right)^{\frac{1}{3}}$, where $H(z_{cl})$ is the Hubble constant computed at the cluster redshift.}. We then computed all profiles as a function of the projected radius in units of $R_{200}$ and rescaled them by the number of members,  $N_0$, found within the radius $R/R_{200}=0.15$ in each cluster. In Fig.~\ref{fig:density2d}, we showed the cumulative projected number and the differential projected number density profiles of cluster members after applying such renormalisations, where the shaded areas correspond to $68\%$ confidence levels. Interestingly, we found that the radial distributions of all clusters followed a universal profile, including \M{0416}, which is an asymmetric merging cluster. We noted that a similar homology relation among rescaled projected mass profiles was found in \citealt{bonamigo2018} and \citealt{caminha2019}, using strong lensing modelling. This result confirms that our methodology was able to identify the \cm{} population with a high degree of purity and completeness.

\section{Conclusions}\label{sec:conc}
In this work, we carry out a detailed analysis of CNN capabilities to identify members in galaxy clusters, disentangling them from foreground and background objects, based on imaging data alone. Such a methodology, therefore, avoided the time consuming and challenging task of building photometric catalogues in cluster cores. We used OPT-NIR high quality HST images, supported by MUSE and CLASH-VLT spectroscopic observations of fifteen clusters, spanning the redshift range $z_{\text{cluster}} = (0.19, 0.60)$.\\ 
We used this extensive spectroscopic coverage to build a training set by combining \cm{s} and \ncm{s}. We performed three experiments by consecutively varying the HST band combinations and the set of training clusters to study the dependence of DL efficiency on (\textit{i}) the cluster redshift  (\textit{EXP1}); and (\textit{ii}) the magnitude and colour of cluster galaxies  (\textit{EXP2}). We also compared the CNN performance with other methods (random forest and Bayesian model), based instead on photometric measurements (\textit{EXP3}).
The main results can be summarised as follows:
\begin{itemize}
    \item[-] Despite members belonging to clusters spanning a wide range of redshift, the CNN achieved a purity-completeness rate $\gtrsim90\%$, showing a stable behaviour and a remarkable generalisation capability over a relatively wide cluster redshift range (Sect.~\ref{ss:exp1}). 
    \item[-] The CNN efficiency was maximised when a large set of sources was combined with HST passbands, including both optical and infrared information. The robustness of the trained model appeared reliable even when a subset of clusters was moved from the training to the blind test set, causing a small drop ($<5\%$) in performance. We observed some performance differences for bright and faint sources, as well as for red and blue galaxies. However, the results maintained the purity, completeness and F1 score greater than 72\% (Table~\ref{tab:exp2:magcolsplit} in Sect.~\ref{ss:exp2}).
    \item[-] By using images, rather than photometric measurements, the CNN technique was able to identify \cm{s} with the lowest rate of contamination and the best trade-off between purity and completeness, when compared to photometry-based methods, which instead require a critical fine-tuning of the classification probability.
    \item[-] The false negatives, that is, the \ncm{s} wrongly classified as \cm{s} were mainly blue and faint. This was simply the result of their limited under-sampling in the training dataset, as well as their similarity with \ncm{s}. 
    However, by excluding sources with $F814>25$mag and $(F606-F814)_{\text{diff}}<-$0.5, the CNN performance improved significantly. These performances reflected the capability of the CNN to classify unknown objects, from which a highly complete and pure magnitude limited sample of candidate \cm{s} could be extracted for several different applications in the study of the galaxy populations and mass distribution of galaxy clusters via lensing techniques. 
\end{itemize}

Therefore, based on an adequate spectroscopic survey of a limited sample of clusters as a training base, the proposed methodology can be considered a valid alternative to photometry-based methods, circumventing the time-consuming process of multi-band photometry, and working directly on multi-band imaging data in counts. To improve CNN performance to recognise the faintest and blue \cm{s}, it would be desirable to plan both HST and spectroscopic observations also covering control fields in the outer cluster regions, with the same depth and passbands as the central regions.

Furthermore, the generalisation capability of this kind of models makes them both versatile and reusable tools. In fact, the convolution layers of a trained deep network can be reused as \textit{shared} layers in larger models, such as the Faster Region CNN \citep{ren2015} and Masked Region CNN \citep{He2017}, which exploit kernel weights to extract multidimensional information suitable to performing  object detection. Such architectures have already found interesting astrophysical applications, for example, in the identification of radio sources \citep{Wu2019} and the automatic deblending of astronomical sources \citep{burke2019}. 

In future works, we will extend this analysis to wide-field ground-based observations and explore other promising deep learning architectures, such as deep auto-encoders \citep{goodfellow:2010} and conditional generative adversarial networks \citep{mehdi2014}, to integrate the ground-based lower resolution images with the high quality of HST images in cluster fields. We also plan to investigate new techniques to overcome the problem of missing data, thus increasing the size of the training set with a more homogeneous sampling of the entire parameter space.


\begin{acknowledgements}
The Authors thank the anonymous referee for the very useful comments and suggestions. The software package HIGHCoOLS\footnote{Hierarchical Generative Hidden Convolution Optimization System, (http://dame.oacn.inaf.it/highcools.html)}, developed within the DAME project \citep{Brescia2014}, has been used for the deep learning models described in this work. We acknowledge funding by PRIN-MIUR 2017WSCC32 “Zooming into dark matter and proto-galaxies with massive lensing clusters”, INAF mainstream 1.05.01.86.20: "Deep and wide view of galaxy clusters (ref. Mario Nonino)" and INAF mainstream 1.05.01.86.31 (ref. Eros Vanzella). MB acknowledges financial contributions from the agreement \textit{ASI/INAF 2018-23-HH.0, Euclid ESA mission - Phase D} and with AM the \textit{INAF PRIN-SKA 2017 program 1.05.01.88.04}. CG acknowledges support through grant no.~10123 of the \textit{VILLUM FONDEN Young Investigator Programme}.
In this work several public software was used: Topcat \citep{Taylor05}, Astropy \citep{astropy2013, astropy2018}, TensorFlow \citep{tensorflow2015}, Keras \citep{keras2015} and Scikit-Learn \citep{sklearn:2011}. 
\end{acknowledgements}

\bibliographystyle{aa}
\bibliography{aanda}

\appendix

\section{Methodology}\label{app:method}
The data preparation phase, preceding the application of the ML based classifiers, is organised as a series of pre-processing steps, detailed in the following sections.

\subsection{Data augmentation}\label{ss:augmentation}
\begin{figure}[htbp]
   \centering
   \includegraphics[width=\columnwidth]{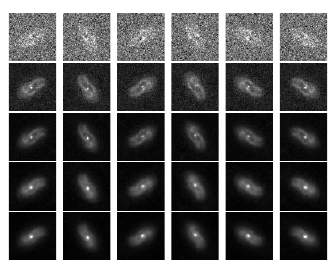}
   \caption{Data augmentation example for a \cm{} at redshift $z=0.531$ (e.g. within the gravitational potential of \M{1149}). Five HST bands are represented from the top to the bottom ($F435$, $F606$, $F814$, $F105$, $F140$). The first column shows the original cut-out, while the three rotations ($\ang{90}$, $\ang{180}$, $\ang{270}$) are reported in columns $2-4$. The two vertical and horizontal flips are shown in the last two columns.}\label{fig:augmented}
\end{figure}
 The cut-outs have been rotated around the three right angles and flipped with respect to the horizontal and vertical axes (an example of such process is shown in Fig.~\ref{fig:augmented}). Given the considerable number of model parameters to fit ($\sim10^5$), deep learning networks require an adequate amount of samples, in order to avoid overfitting \citep{Cui2015, Perez2017}. However, an uncontrolled augmentation could introduce false correlations among the training samples. Therefore, only a fraction of sources have been subject to these transformations: $15\%$ of the available images have been randomly extracted and used for such transformations mentioned above. The resulting augmentation factor was $1.75$ times the original dimension of the training set. Obviously, such augmentation process involved only the training images. 

\subsection{Setup of training and test sets}\label{ss:traintest}

\begin{figure*}[htbp]
   \centering
   \includegraphics[width=2\columnwidth]{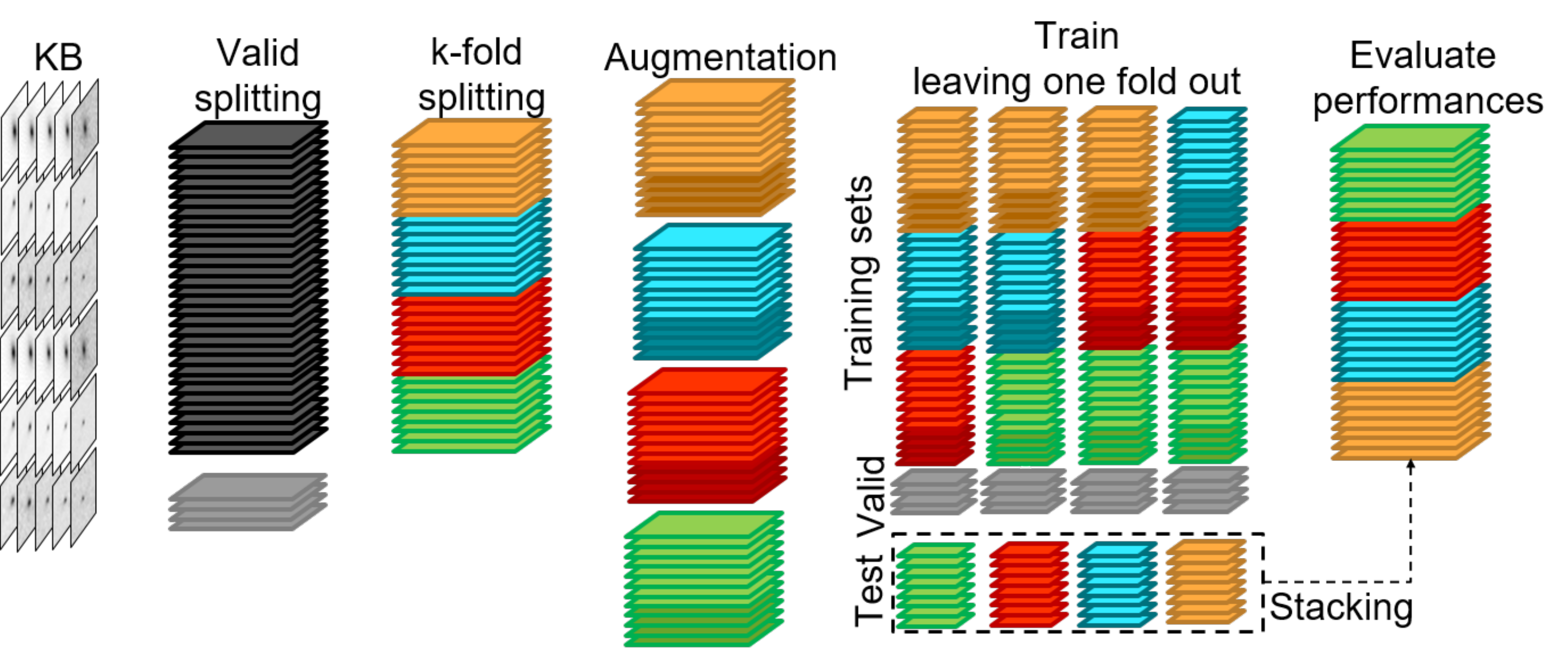}
   \caption{Data preparation flow: from the whole dataset (i.e. the knowledge base) a validation set is extracted. The rest of the dataset is split through a k-fold partitioning process (in this image, we simplified the figure assuming $k=4$ folds, while in reality we used $k=10$). The training samples are then arranged, by permuting the involved augmented folds, while the test samples do not include the artefact images generated by the augmentation process. These sets are finally stacked in order to evaluate the global training performances.}\label{fig:dataflow}
\end{figure*}

Before the k-fold splitting and the augmentation process, described above, we randomly extracted a small sample of sources ($10\%$ of the data set), reserved as validation set during the training phase in order to control the gradual reduction of the learning rate on the plateau of the cost function \citep{bengio:2012} and an early stopping regularisation process \citep{Prechelt1997, Raskutti:2011}.  
The data preparation flow is depicted in Fig.~\ref{fig:dataflow}: (\textit{i}) the dataset is composed by multi-bands images; (\textit{ii}) a fraction of sources ($10\%$) is extracted as validation set; (\textit{iii}) the remaining samples are split into $k=10$ folds without overlapping; (\textit{iv}) for each of them, a fraction ($15\%$) of samples is augmented through cut-out rotations and flips; (\textit{v}) the training sets are built by concatenating $k-1$ folds (composed by the original images and the artefacts) and the learning is evaluated on the $k$-th fold (without artefacts), acting as blind test; (\textit{vi}) finally, the model performances are evaluated on the whole training set, obtained by stacking all its (test) folds.

\subsection{Statistical evaluation of performance}\label{ss:estimators}
In order to assess the model classification performances, we chose the following statistical estimators: `average efficiency' (among all classes, abbreviated as `AE'), `purity' (also know as `positive predictive value' or 'precision', abbreviated as `pur'), `completeness' (also known as `true positive rate' or `recall', abbreviated as `comp'), and F1-score (a measure of the combination of purity and completeness, abbreviated as `F1'). 

\begin{table}[htbp]\caption{Generic confusion matrix for a binary classification problem.}\label{tab:cmexp}\centering\small
\begin{tabular}{ccccc}
\hline
&&\multicolumn{2}{c}{Predictions} \\
&&positive & negative\\\hline
\multirow{2}{*}{True}&positive & \textbf{TP}& \textbf{FN}\\\cline{2-4}
&negative & \textbf{FP} & \textbf{TN}\\\hline
\end{tabular}
\tablefoot{In a confusion matrix, columns indicate the number of objects per class, as predicted by the classifier, while rows are referred to the true (known) objects per class. Hence, the main diagonal terms report the number of correctly classified objects for each class. While, the terms FP and FN count, respectively, the false positives and false negative quantities.}
\end{table}

In a binary confusion matrix, as in the example shown in Table~\ref{tab:cmexp}, columns indicate the class objects as predicted by the classifier, while rows refer to the true objects per class. The main diagonal terms contain the number of correctly classified objects for each class, while the terms FP and FN report the amount of, respectively, false positives and false negatives. Therefore, the derived estimators are computed as:
\begin{eqnarray}
\text{AE} & = & \frac{TP+TN}{TP+FP+TN+FN}\\
\text{pur} & = & \frac{TP}{TP+FP}\\
\text{comp} & = & \frac{TP}{TP+FN}\\
\text{F1} & = & 2\cdot\frac{\text{pur}\cdot\text{comp}}{\text{pur}+\text{comp}}
\end{eqnarray}

The \textit{AE} is the ratio between the sum of the correctly classified objects (for all the involved classes) and the total amount of objects; it describes an average evaluation weighted on all involved classes. The `purity' of a class is the ratio between the correctly classified objects and the sum of all objects assigned to that class (i.e. the predicted membership); it measures the precision of the classification. The `completeness' of a class is the ratio between the correctly classified objects and the total amount of objects belonging to that class (i.e. the `true' membership), it estimates the sensitivity of the classification. Finally, the \textit{F1-score} is the harmonic average between purity and completeness.
By definition, the dual quantity of purity is the `contamination', a measure which indicates the amount of misclassified objects for each class.  

Moreover, from the probability vector (i.e. the set of values stating the probability that an input belongs to a certain class), it is possible to extract another useful estimator, the receiver operating characteristic (ROC) curve. It is a diagram in which the true positive rate is plotted versus the false positive rate by varying a membership probability threshold (see Fig.~\ref{fig:exp2}). The overall score is measured by the area under the ROC curve (AUC), where an area of $1$ represents a perfect classification, while an area of $0.5$ indicates a useless result (akin to a toss of a coin). 

\section{Convolution Neural Networks}\label{app:CNN}
\begin{figure*}[htbp]
   \centering
   \includegraphics[width=2\columnwidth]{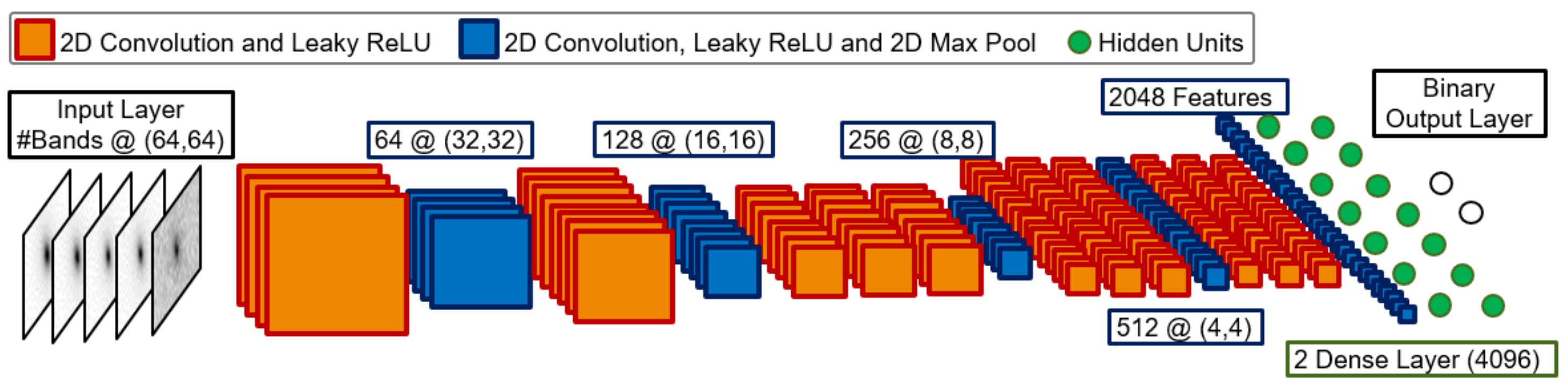}
   \caption{Streamlined representation of the architecture designed for the CNN model used in this work. Orange and blue items describe two different block operations, respectively: (i) convolution and activation function, (ii) convolution, activation function and pooling. The simultaneous reduction of the square dimensions and their increasing amount intuitively represent the abstraction process typical of a CNN. Green circular units are arranged in order to describe the fully connected (i.e. dense) layers. The dimensions of the feature maps are reported for each pooling operation, together with the number of features extracted by the CNN.}\label{fig:cnn}
\end{figure*}

In this appendix, CNNs theory and our specific implementation are briefly described. while a synthetic view of the implemented model is shown in Fig.~\ref{fig:cnn}.

As any other artificial neural networks, convolution neural networks \citep[CNNs,][]{LeCun:1989} are inspired by biological behaviours. Artificial neurons are arranged in several layers, where each neuron takes as input the signal coming from neurons belonging to the previous layer; such as biological neurons, the variation of the synaptic connection sensibility (with respect to a certain input signal) is correlated to the learning mechanism \citep{hebb:1949}. During the training, these connection sensibilities among layers (i.e. the weights) are adapted through a forward-backward mechanism, at the base of the iterative learning process \citep{Bishop:2006}. After training, supervised Machine Learning methods define a non-linear relation between the input and output spaces, which is encoded within the weight matrices. 

CNNs represent one of the most widely-used supervised techniques among the Deep Neural Networks (DNN, \citealt{goodfellow:2016}), whose peculiarity is an ensemble of receptive fields which trigger neuron activity. The receptive field is represented by a small matrix (called as kernel or filter), which connects two consecutive layers through a convolution operation. Similar to the adaptation mechanism imposed by supervised machine learning, the kernels are modified during the training. The peculiarity of such kind of models is the capability to automatically extract meaningful features from images (such as edges and shapes), which become the input vector to any standard ML model that outputs the class of the input image. The idea behind CNN is a convolution-subsampling chain mechanism: deep networks are characterised by tens of layers (in some cases hundreds, as proposed by \citealt{he2015} and \citealt{Xie:2016}), where at each depth level, the convolution acts as a filter, emphasising (or suppressing) some properties; while the subsampling (often called pooling) makes sure that only essential information is moved towards the next layer.

CNNs are organised as a hierarchical series of layers, typically based on convolution and pooling operations. Convolution kernel is represented by a $4$-D matrix $\mathbf{K}$, where the element $K_{i,j,k,l}$ is the connection weight between the output unit $i$ and the input unit $j$, with an offset of $k$ rows and $l$ columns. This kernel is convoluted with the input signal and adapted during the training. Given an input $\mathbf{V}$, whose element $V_{i,j,k}$ represents an observed data value of the channel $i$ at row $j$ and column $k$, the neuron activity can be expressed as \citep{goodfellow:2010}:
\begin{eqnarray}
        Z_{i,j,k} & = & c(\mathbf{K}, \mathbf{V}, s)_{i,j,k} + b = \nonumber\\
        & = & \sum_{l,m,n} V_{l, (j-1)\times s+m, (k-1)\times s +n}K_{i, l, m, n} + b \\
        Z_{i,j,k} & \xleftarrow{} & p(\mathbf{Z}, d)_{i,j,k}\\
        Z_{i,j,k} & \xleftarrow{} & f(\mathbf{Z}, \{a\}_q)_{i,j,k}
\end{eqnarray}
where $c(\mathbf{K}, \mathbf{V}, s)$ is the convolution operation between the input $\mathbf{V}$ and the kernel $\mathbf{K}$ with stride $s$; $b$ is an addend that acts as bias; $p(\mathbf{Z}, d)$ is the pooling operation with down-sampling factor $d$; $f(\mathbf{Z}, \{a\}_q)$ is the activation function characterised by the set of hyper-parameters $\{a\}_q$. 
The pooling function reduces the dimension, by replacing the network output at a certain location with a summary statistic of nearby outputs \citep{goodfellow:2016}. 

Unlike traditional artificial neural networks (e.g. Multi-Layer Perceptron), where all neurons of two consecutive layers are fully connected among them, the connection among neurons in a CNN is `sparse', that is, the interaction between neurons belonging to different layers is limited to a small fraction. This reduces the number of operations, the memory requirements and, thus, the computing time. The output layer consists of an ensemble of sub-images with reduced dimensions, called feature maps, each of them represents a feature extracted from the original signal, processed by the net in order to solve the assigned problem. 

\begin{table}[htbp]\caption{VGGNET-like model configuration.} \label{tab:nethyperparams}\centering\small
\begin{tabular}{P{2.0cm}P{2.0cm}P{1.7cm}}
\hline
Layer & Output Shape & Params \# \\\hline
Input layer & (64, 64, NC) & 0 \\\hline
Conv2D & (64, 64, 64) & 6976 \\\hline
Leaky ReLU & (64, 64, 64) & 0 \\\hline
Conv2D & (64, 64, 64) & 36928 \\\hline
Leaky ReLU & (64, 64, 64) & 0 \\\hline
Max Pool2D & (32, 32, 64) & 0 \\\hline
Conv2D & (32, 32, 128) & 73856 \\\hline
Leaky ReLU & (32, 32, 128) & 0 \\\hline
Conv2D & (32, 32, 128) & 147584 \\\hline
Leaky ReLU & (32, 32, 128) & 0 \\\hline
Max Pool2D & (16, 16, 128) & 0 \\\hline
Conv2D & (16, 16, 256) & 295168 \\\hline
Leaky ReLU & (16, 32, 256) & 0 \\\hline
Conv2D & (16, 16, 256) & 590080 \\\hline
Leaky ReLU & (16, 16, 256) & 0 \\\hline
Conv2D & (16, 16, 256) & 590080 \\\hline
Leaky ReLU & (16, 32, 256) & 0 \\\hline
Conv2D & (16, 16, 256) & 590080 \\\hline
Leaky ReLU & (16, 16, 256) & 0 \\\hline
Max Pool2D & (8, 8, 256) & 0 \\\hline
Conv2D & (8, 8, 512) & 1180160 \\\hline
Leaky ReLU & (8, 8, 512) & 0 \\\hline
Conv2D & (8, 8, 512) & 2359808 \\\hline
Leaky ReLU & (8, 8, 512) & 0 \\\hline
Conv2D & (8, 8, 512) & 2359808 \\\hline
Leaky ReLU & (8, 8, 512) & 0 \\\hline
Conv2D & (8, 8, 512) & 2359808 \\\hline
Leaky ReLU & (8, 8, 512) & 0 \\\hline
Max Pool2D & (4, 4, 512) & 0 \\\hline
Conv2D & (4, 4, 512) & 2359808 \\\hline
Leaky ReLU & (4, 4, 512) & 0 \\\hline
Conv2D & (8, 8, 512) & 2359808 \\\hline
Leaky ReLU & (4, 4, 512) & 0 \\\hline
Conv2D & (4, 4, 512) & 2359808 \\\hline
Leaky ReLU & (4, 4, 512) & 0 \\\hline
Conv2D & (4, 4, 512) & 2359808 \\\hline
Leaky ReLU & (4, 4, 512) & 0 \\\hline
Max Pool2D & (2, 2, 512) & 0 \\\hline
Flatten & (2048) & 0 \\\hline
Dense &  (4096) & 8392704\\\hline
Leaky ReLU & (4096) & 0 \\\hline
Dropout & (4096) & 0 \\\hline
Dense &  (4096) & 16781312  \\\hline
Leaky ReLU & (4096) & 0 \\\hline
Dropout & (4096) & 0 \\\hline
Dense &  (2) & 8194  \\\hline
Output Layer & (2) & 0 \\\hline
\end{tabular}
\tablefoot{The columns specify the layer operation, the shape of the output and the number of parameters to fit. The output shape of a layer is a $4$-D matrix, but, since the first dimension is the fixed size of the input data batch (with a size of $64$ patterns), we do not mention this number to prevent confusion. The total amount of trainable parameters is larger than $45$M. The last dimension of the input layer is the involved number of channels (i.e. the number of photometric bands used), a quantity depending on the specific experiment (see Sect.~\ref{sec:data}).}
\end{table}

Another common operation performed during the training of a CNN is the random dropout of weights. This function prevents units from co-adapting, reduces significantly overfitting and gives major improvements over other regularization methods \citep{srivastava:2014}. At the end of the network, the resulting feature maps ensemble is fully connected with one or more hidden layers (also called `dense layers'), the last of which, in turn, is fully connected to the output layer. The net output must have the same shape of the known target: within the supervised learning paradigm, the comparison between output and target induces the kernel adaptations. When the net task is a classification problem (as in this work), the output is a matrix of probabilities, that is, each sample has a membership probability related to any class of the problem. In order to transform floating values into probabilities (i.e. forced to the constraint $\sum_{j=1}^{n_{classes}} p_j =1$), the activation function of the final dense layer is typically a \textit{softmax}, which normalises a vector into a probability distribution \citep{Bishop:2006}. In order to solve a classification problem, the network learns how to disentangle objects in the train set, minimising a loss function (or cost function). The most common choice for the loss function is the \textit{cross-entropy} \citep{goodfellow:2016}:
\begin{equation}
    C(y,\bar{y}) = \sum_{j=1}^{n_{classes}}y_j\ln{\bar{y}_j}+(1-y_j)\ln{(1-\bar{y}_j)}
\end{equation}
where $y$ is the target and $\bar{y}$ is the output of the final layer. Thus, during the training, images extracted from the train set are propagated through the network, while weights and biases are adapted along with a backward flow in order to minimise the cost function. To perform such optimization, it is necessary to state the minimization algorithm. The simplest and most used optimiser is the Stochastic Gradient Descendent \citep{Bishop:2006}, but in recent years several optimisers have been proposed \citep[e.g.][]{Duchi:2011,Zeiler:2012,Diederik:2018}, which offer a faster convergence to the minimum, avoiding the local dump of the cost function. In this work we chose \textit{Adadelta} \citep{Zeiler:2012} as optimiser. Furthermore, we included \textit{(i)} an \textit{early stopping} regularisation criterion \citep{Prechelt1997, Raskutti:2011}, preventing overfitting; and \textit{(ii)} a gradual reduction of the learning rate on the plateau of the loss function \citep[as function of epochs,][]{bengio:2012}. Both techniques exploit a validation set, extracted from the train set, used to compute and evaluate the learning efficiency within the training cycles. In our case, to avoid memory loss, the network has been trained with input data batches of size equals to $64$ patterns.

The architecture of our VGGNET-like model is reported in Table~\ref{tab:nethyperparams}. It is composed of $47$ layers and convolution kernels with a window size of $3\times 3$. The \textit{max} pooling criterion was preferred to the \textit{average} algorithm, in order to reduce the noise contribution. We set the Leaky version of a Rectified Linear Unit \citep[LeReLU,][]{Maas:2013} as activation function for all the neurons. This type of activation allows \textit{(i)} a small, non-zero gradient also when the unit is saturated and not active, \textit{(ii)} a gain of the convergence with the increase of the units, defined as:
\begin{equation}
    out(x)=\begin{dcases}
	   alpha \cdot x & \quad x< 0\\
	   x  & \quad x \ge 0\\
    \end{dcases}
\end{equation}
where $\alpha$ is a hyper-parameter set to $0.3$.  

This network has been implemented through \textit{keras} \citep{keras2015}, with \textit{tensorflow} \citep{tensorflow2015} as backend system. Both of them are open-source Python libraries, allowing the automatic handling of the Graphic Processing Unit (GPU), achieving a huge gain in terms of computational cost \citep[$\sim 700$ see][]{simard:2005}. In this work the experiments were performed with an NVIDIA GPU Titan Xp and an NVIDIA GPU Quadro P5000, requiring $\sim 30$ minutes to complete the training (on a single fold, see Sect.~\ref{ss:traintest}).

\section{Benchmark methods}\label{app:benchmethods}
We compared CNN performances with two techniques based on photometric catalogues: a random forest (RF, \citealt{breiman:2001}) and a Bayesian Method (briefly described in \citealt{grillo2015}). 

A Bayesian classifier is a model able to minimise the error probability \citep{devroye1996}, defined as: $L(g)=P[g(X)\ne Y]$, where $(X, Y)$ are pair values $\in \mathbb{R}^d\times\{1, \dots, M\}$ (i.e. $Y$ is the ensemble of class labels related to the manifold $X$), $g$ is a classifier (i.e. a function $g: x\in X \subseteq \mathbb{R}^d\rightarrow y\in \{1, \dots, M\}$), $L$ is an application mapping $g$ into probabilities. The minimal probability error is denoted $L^*=L(g^*)$, that can be written as: 
$$g^* = \underset{g:\mathbb{R}^d\rightarrow \{1, \dots, M\}}{\text{argmin}} P[g(X)\ne Y] $$
Given a classical linear model $\bar{y}_i=\sum_{j=1}^p x_{ij}\theta_j $, $ i=1, \dots, n$, the method estimates $\{\theta\}_i^p$ in order to minimise a coherent combination of the residuals $r_i = y_i-\bar{y}_i$. The implemented method exploits a minimum covariance determinant method \citep{rousseeuw1984}, which is based on the minimization of the median of squared residuals.

Random forest is a machine learning classifier consisting of a collection of tree-structured classifiers $\{h(x,\Theta_k), k=1, ...\}$ where the $\{\Theta_k \}$ are independent identically distributed random vectors and each tree casts a unit vote for the most popular class at input x. The generalisation error for this algorithm depends on the strength of single trees and from their correlations through the raw margin functions. To improve the model accuracy by keeping trees strength, the correlation between trees is decreased and bagging with a random selection of features is adopted. Bagging, or bootstrap aggregating, is a method designed to improve the stability and accuracy of machine learning algorithms. It also reduces variance and helps to avoid overfitting. In this work, we used the RF provided by Scikit-Learn python library \citep{sklearn:2011}.

\section{Technical descriptions of our performed experiments}\label{app:experiments}
In this appendix, we report tables and figures describing in detail the analysis performed for each experiment.

\subsection{EXP1}\label{app:Exp1}
With this experiment, we evaluated the CNN capabilities to identify \cm{s} at different cluster redshifts, $z_{\text{cluster}} \in (0.2, 0.6)$, using different HST band combinations (see Sections~\ref{sec:data} and \ref{ss:exp1}). Furthermore, in this experiment, we studied the dependence on redshift and on the number of spectroscopic sources involved in the training. The results related to this experiment have been summarised in Sect.~\ref{ss:exp1}. 

In Table~\ref{tab:exp1:stack}, we report the results achieved globally by CNN, that is, by combining the available clusters (see also Fig.~\ref{fig:exp1:stack}), while Table~\ref{tab:exp1:clusters} outlines the performances for each involved cluster, varying the band combinations. The experiment has been carried out with the k-fold approach, stacking sources in the FoV of 13 (15 only for the \textit{mixed*} configuration) clusters, ensuring that the k-est fold is populated by objects extracted from each involved cluster, proportionally to the number of available spectroscopic sources, that is, providing adequate coverage of the training set respect to the test set.

\begin{figure*}[htbp]
   \centering
   \includegraphics[width=2\columnwidth]{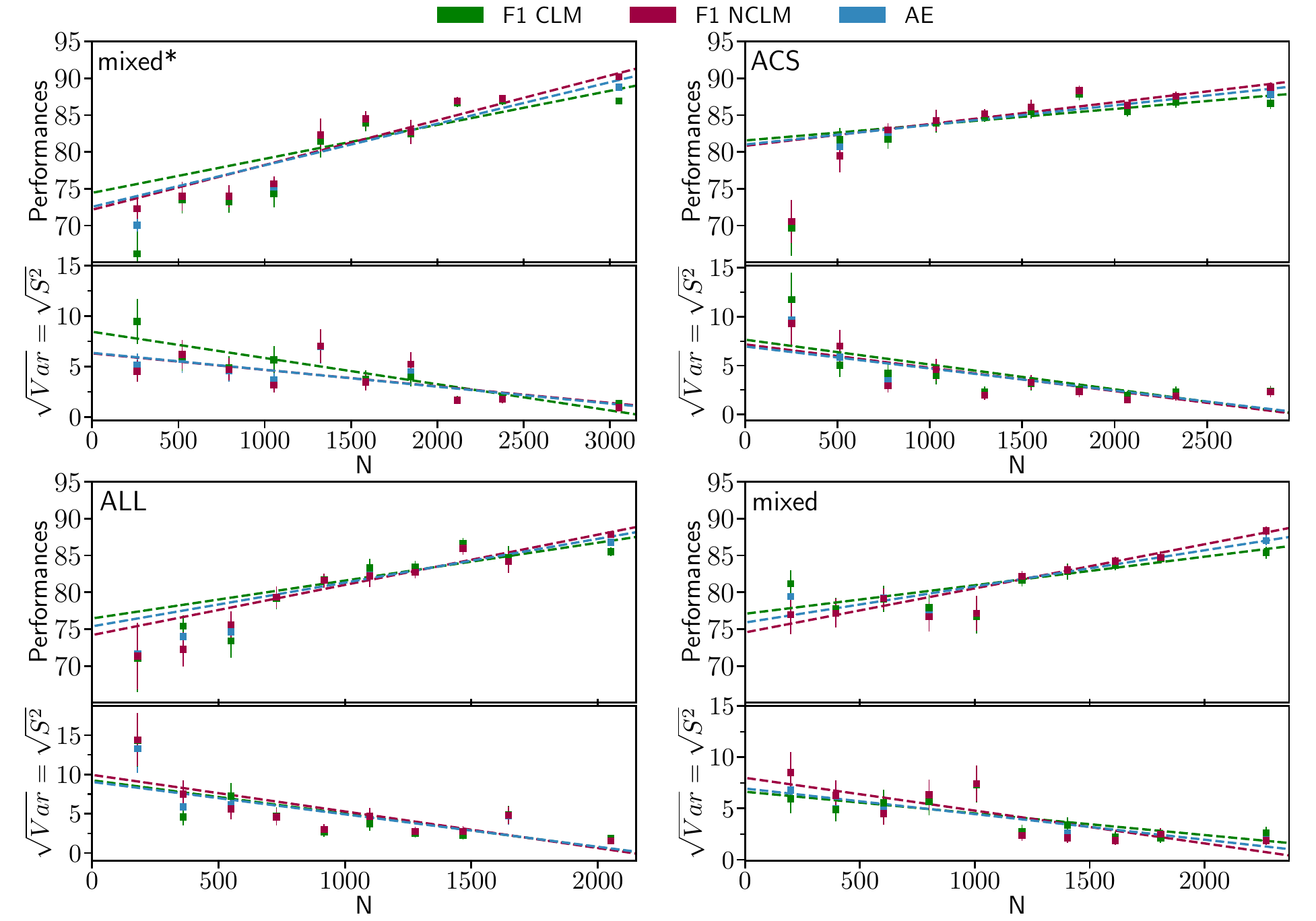}
   \caption{Comparison among the four band configurations (see Sect.~\ref{sec:data}), in terms of F1 score and average efficiency (AE) percentages (top panels), together with their square root of variances (bottom panels), as the number of spectroscopic sources in the training set increases (\textit{EXP1}). In all panels\textbf{,} the linear best\textbf{-}fit trends are displayed as dashed lines. Due to the k-fold approach, performances have been averaged over the $10$ folds, i.e. the x-axis shows the dimension of the training set, thus, the k-est fold used as test set has a size of $N/9$.}\label{fig:exp1:sizedep}
\end{figure*}

The comparison between the band configurations is also shown in Fig.~\ref{fig:exp1:sizedep}, in which performances and their fluctuations are displayed as function of the involved number of samples. For each configuration, we split the knowledge space into ten disjointed subsets, which have been progressively merged in order to build a dataset with which CNN has been trained and tested, always using the k-fold approach. 

In order to analyse the dependence on redshift, we split the \cm{} redshift range into five equally populated bins and, to complete the knowledge space, with extracted without repetitions from the \ncm{} population an appropriate number of objects. The network has been trained within each ensemble adopting the k-fold approach, using only the \textit{mixed*} band combination. The result is graphically shown in Fig.~\ref{fig:exp1:zbin} and it is stored in Table~\ref{tab:exp1:zbin}, in which, we have specified the fluctuation of estimators as an error estimated on the ten folds. 

\begin{table}[htbp]\caption{CNN percentage performances in the \textit{EXP1} experiment} \label{tab:exp1:stack}\centering
\begin{tabular}{llcccc}
\hline
\textit{Class} & \% & \textit{mixed} & \textit{ACS} & \textit{ALL} & \textit{mixed*} \\\hline
& \textit{AE} & 86.7 & 87.4 & 87.7 & \textbf{89.3} \\\hline
& \textit{pur} & 83.1 & 85.0 & 86.4 & \textbf{88.3} \\
\cm{} & \textit{compl} & 88.4 & \textbf{88.5} & 86.4 & 86.7 \\
& \textit{F1} & 85.6 & 86.7 & 86.4 & \textbf{87.4} \\\hline
& \textit{pur} & \textbf{90.0} & 89.9 & 88.9 & \textbf{90.0} \\
\ncm{} & \textit{compl} & 85.5 & 86.7 & 88.9 & \textbf{91.2} \\
& \textit{F1} & 87.7 & 88.3 & 88.9 & \textbf{90.6} 
\end{tabular}
\tablefoot{The performances are related to the four band configurations (see Sect.~\ref{sec:data}) and expressed in terms of the statistical estimators described in Sect.~\ref{ss:estimators}. The overall best results are highlighted in bold.}
\end{table}

\begin{table*}[htbp]\caption{CNN percentage performances evaluated for each cluster and for each band configuration related to the \textit{EXP1} experiment.}\label{tab:exp1:clusters}\centering
\begin{tabular}{llcccccccccccc}\\\hline
&&\multicolumn{4}{c}{\A{383} $z=\zcluster{a383}$} &\multicolumn{4}{c}{\R{2129} $z=\zcluster{r2129}$} &\multicolumn{4}{c}{\A{2744} $z=\zcluster{a2744}$}\\\hline 
&& mixed & ACS & ALL & mixed* & mixed & ACS & ALL & mixed* & mixed & ACS & ALL & mixed*\\
&AE & 77.0 & 81.8 & 78.3 & \textbf{83.0} & 89.7 & 91.6 & \textbf{93.7} & 92.3 & & & & 93.6 \\
&pur & 77.2 & 82.9 & 82.5 & \textbf{86.3} & 76.5 & 84.6 & \textbf{86.5} & 84.4 &&&& 95.3 \\
\cm{} &compl & 81.5 & \textbf{82.9} & 75.0 & 81.5  & 90.7 & \textbf{91.7} & 88.9 & 88.4  &&&& 86.5 \\
&F1 & 79.3 & 82.9 & 78.6 & \textbf{83.8} & 83.0 & \textbf{88.0} & 87.7 & 86.4  &\multicolumn{3}{c}{only mixed*}& 90.7 \\
&pur & 76.7 & \textbf{80.6} & 74.4 & 79.6  & \textbf{96.2} & 95.6 & \textbf{96.2} & 95.5  &&&& 92.8 \\
\ncm{} &compl & 71.7 & 80.6 & 82.1 & \textbf{84.8}  & 89.3 & 91.6 & \textbf{95.3} & 93.8  &&&& 97.6 \\
&F1 & 74.2 & 80.6 & 78.0 & \textbf{82.1}  & 92.6 & 93.6 & \textbf{95.8} & 94.6  &&&& 95.2 \\\hline

&&\multicolumn{4}{c}{\MS{2137} $z=\zcluster{m2137}$}&\multicolumn{4}{c}{\R{2248} $z=\zcluster{r2248}$} &\multicolumn{4}{c}{\M{1931} $z=\zcluster{m1931}$}\\\hline 
&& mixed & ACS & ALL & mixed* & mixed & ACS & ALL & mixed* & mixed & ACS & ALL & mixed*\\
&AE & 83.7 & 81.5 & 88.2 & \textbf{88.4} & 89.5 & 86.5 & \textbf{90.2} & 88.1 & 84.0 & 86.0 & 84.9 & \textbf{90.0} \\
&pur & 80.0 & 79.7 & 85.7 & \textbf{89.7} & 88.6 & 85.2 & \textbf{90.7} & 88.3 & 91.3 & 85.3 & 83.6 & \textbf{100.0} \\
\cm{} &compl & 87.8 & 81.0 & \textbf{90.9} & 85.4 & 92.4 & 91.3 & \textbf{92.5} & 89.8 & 67.7 & \textbf{80.6} & 78.0 & 75.8 \\
&F1 & 83.7 & 80.3 & \textbf{88.2} & 87.5 & 90.5 & 88.1 & \textbf{91.6} & 89.1 & 77.8 & 82.9 & 80.7 & \textbf{86.2} \\
&pur & 87.8 & 83.1 & \textbf{90.9} & 87.2 & \textbf{90.6} & 88.3 & 89.5 & 87.9 & 80.8 & \textbf{86.4} & 85.7 & 85.4 \\
\ncm{} &compl & 80.0 & 81.9 & 85.7 & \textbf{91.1} & 86.1 & 80.7 & \textbf{87.2} & 86.1 & 95.5 & 89.9 & 89.7 & \textbf{100.0} \\
&F1 & 83.7 & 82.5 & 88.2 & \textbf{89.1} & \textbf{88.3} & 84.3 & 88.3 & 87.0 & 87.5 & 88.1 & 87.6 & \textbf{92.1} \\\hline

&&\multicolumn{4}{c}{\M{1115} $z=\zcluster{m1115}$} & \multicolumn{4}{c}{\A{370} $z=\zcluster{a370}$} &\multicolumn{4}{c}{\M{0416} $z=\zcluster{m0416}$}\\\hline 
&& mixed & ACS & ALL & mixed* & mixed & ACS & ALL & mixed* & mixed & ACS & ALL & mixed*\\
&AE & 88.1 & 84.9 & 89.6 & \textbf{92.5}  &&&& 88.9  & 90.3 & 90.0 & 91.5 & \textbf{92.2} \\
&pur & 85.7 & 82.5 & 90.9 & \textbf{91.8}  &&&& 85.8  & 92.4 & 90.3 & \textbf{95.7} & 93.3 \\
\cm{} &compl & 93.0 & 89.5 & 89.3 & \textbf{94.4}  &&&& 87.6  & 87.1 & \textbf{88.8} & 86.8 & 87.1 \\
&F1 & 89.2 & 85.8 & 90.1 & \textbf{93.1}  &\multicolumn{3}{c}{only mixed*}& 86.7  & 89.7 & 89.5 & 91.0 & \textbf{91.5} \\
&pur & 91.2 & 87.9 & 88.2 & \textbf{93.4}  &&&& 89.5  & 88.6 & 89.7 & 88.1 & \textbf{89.0} \\
\ncm{} &compl & 82.5 & 80.0 & 90.0 & \textbf{90.5}  &&&& 87.9  & 93.3 & 91.1 & 96.1 & \textbf{96.9} \\
&F1 & 86.7 & 83.8 & 89.1 & \textbf{91.9}  &&&& 88.6  & 90.9 & 90.4 & 91.9 & \textbf{92.8} \\\hline

&&\multicolumn{4}{c}{\M{1206} $z=\zcluster{m1206}$}&\multicolumn{4}{c}{\M{0329} $z=\zcluster{m0329}$} &\multicolumn{4}{c}{\R{1347} $z=\zcluster{r1347}$}\\\hline 
&& mixed & ACS & ALL & mixed* & mixed & ACS & ALL & mixed* & mixed & ACS & ALL & mixed*\\
&AE & 87.7 & \textbf{90.3} & 87.4 & 89.7 & 81.6 & 81.9 & 83.3 & \textbf{85.0}  & \textbf{91.2} & 90.7 & 89.7 & 89.9 \\
&pur & 83.7 & 89.8 & 84.2 & \textbf{89.9} & 76.9 & 76.8 & 79.1 & \textbf{83.3} & 79.7 & \textbf{81.6} & 80.4 & 81.0 \\
\cm{} &compl & 89.7 & \textbf{90.2} & 86.7 & 86.5 & 89.6 & \textbf{91.5} & 88.3 & 91.0 & \textbf{100.0} & 96.9 & 93.8 & 92.2 \\
&F1 & 86.6 & \textbf{90.0} & 85.4 & 88.2 & 82.8 & 83.5 & 83.5 & \textbf{87.1} & \textbf{88.7} & 88.6 & 86.5 & 86.2 \\
&pur & \textbf{91.3} & 90.8 & 89.9 & 89.6 & 87.9 & 89.5 & 88.1 & \textbf{90.0} & \textbf{100.0} & 97.9 & 96.2 & 95.6 \\
\ncm{} &compl & 86.2 & 90.4 & 88.0 & \textbf{92.3} & 73.9 & 72.3 &\textbf{ 78.8} & 78.3  & 86.6 & 87.0 & 87.5 & \textbf{88.7} \\
&F1 & 88.7 & 90.6 & 88.9 & \textbf{90.9} & 80.3 & 80.0 & 83.2 & \textbf{83.7} & \textbf{92.8} & 92.2 & 91.7 & 92.0 \\\hline

&&\multicolumn{4}{c}{\M{1311} $z=\zcluster{m1311}$} &\multicolumn{4}{c}{\M{1149} $z=\zcluster{m1149}$} &\multicolumn{4}{c}{\M{2129} $z=\zcluster{m2129}$}\\\hline 
&& mixed & ACS & ALL & mixed* & mixed & ACS & ALL & mixed* & mixed & ACS & ALL & mixed*\\
&AE & 77.1 & \textbf{82.5} & 75.8 & 78.1  & 85.9 & \textbf{90.7} & 88.0 & 89.4 & 85.5 & \textbf{86.4} & 84.9 & 86.1 \\
&pur & 72.7 & \textbf{80.3} & 75.0 & 76.0  & 74.5 & \textbf{83.3} & 80.5 & 82.3 & 85.9 & 87.3 & 91.0 & \textbf{91.3} \\
\cm{} &compl & \textbf{85.1} & 77.8 & 78.3 & 80.9  & \textbf{94.5} & 92.6 & 91.5 & 91.3 & 82.7 & \textbf{83.1} & 75.3 & 77.8 \\
&F1 & 78.4 & \textbf{79.0} & 76.6 & 78.4  & 83.3 & \textbf{87.7} & 85.6 & 86.6 & 84.3 & \textbf{85.2} & 82.4 & 84.0 \\
&pur & 82.9 & \textbf{84.1} & 76.7 & 80.4  & \textbf{96.1} & 95.6 & 94.0 & 94.5 & 85.3 & \textbf{85.6} & 81.0 & 82.7 \\
\ncm{} &compl & 69.4 & \textbf{86.0} & 73.3 & 75.5  & 80.8 & \textbf{89.7} & 85.7 & 88.3 & 88.0 & 89.2 & 93.4 & \textbf{93.5} \\
&F1 & 75.6 & \textbf{85.1} & 75.0 & 77.9  & 87.8 & \textbf{92.6} & 89.7 & 91.3 & 86.6 & 87.4 & 86.7 & \textbf{87.8}
\end{tabular}
\end{table*}

\begin{table*}[htbp]\caption{Statistical estimators measured in each redshift bin for the \textit{EXP1a} experiment.}\label{tab:exp1:zbin}
\resizebox{2\columnwidth}{!}{
\begin{tabular}{llcc cc cc cc cc}\\\hline
&& k-fold & global & k-fold & global & k-fold & global & k-fold & global & k-fold & global \\\hline
& & \multicolumn{2}{l}{z$_{\cm{}}$ $\in$ (0.18, 0.32)} & \multicolumn{2}{l}{z$_{\cm{}}$ $\in$ (0.32, 0.37)} & \multicolumn{2}{l}{z$_{\cm{}}$ $\in$ (0.37, 0.41)} & \multicolumn{2}{l}{z$_{\cm{}}$ $\in$ (0.41, 0.46)} & \multicolumn{2}{l}{z$_{\cm{}}$ $\in$ (0.46, 0.60)}\\\hline
& AE & $86.4\pm1.1$ & 86.2 & $89.0\pm1.2$ & 89.2 & $88.8\pm1.4$ & 88.6 & $88.1\pm1.0$ & 87.9 & $89.6\pm1.3$ & 89.6 \\
& pur & $84.9\pm2.6$ & 84.1 & $87.0\pm1.7$ & 86.9 & $87.9\pm1.6$ & 87.3 & $87.1\pm1.0$ & 87.0 & $87.7\pm2.0$ & 87.3 \\
\ncm{} & compl & $89.6\pm1.6$ & 89.2 & $92.1\pm1.8$ & 92.4 & $90.3\pm1.5$ & 90.3 & $89.5\pm1.6$ & 89.2 & $92.8\pm0.9$ & 92.7 \\
& F1 & $86.9\pm0.9$ & 86.6 & $89.3\pm1.2$ & 89.5 & $89.0\pm1.3$ & 88.8 & $88.2\pm1.0$ & 88.1 & $90.0\pm1.2$ & 89.9 \\
& pur & $89.3\pm1.1$ & 88.5 & $91.7\pm1.7$ & 91.8 & $90.2\pm1.5$ & 89.9 & $89.4\pm1.5$ & 88.9 & $92.5\pm0.9$ & 92.2 \\
\cm{} & compl & $83.1\pm3.2$ & 83.1 & $85.9\pm2.1$ & 86.0 & $87.3\pm1.8$ & 86.9 & $86.6\pm1.2$ & 86.6 & $86.5\pm2.2$ & 86.5 \\
& F1 & $85.7\pm1.4$ & 85.7  & $88.5\pm1.3$ & 88.9  & $88.6\pm1.4$ & 88.4  & $87.9\pm1.0$ & 87.8  & $89.2\pm1.4$ & 89.2 \\
\end{tabular}}
\tablefoot{Due to the k-fold approach, the performances are reported as pairs of mean and error (evaluated on the $10$ folds) and as a single global value.}
\end{table*}

\subsection{EXP2}\label{app:Exp2}
In this experiment, we explored the limits of the CNN in terms of its classification efficiency. With this aim, we excluded three clusters from the training sample, respectively, \A{370} ($z=\zcluster{a370}$), \MS{2137} ($z=\zcluster{m2137}$) and \M{0329} ($z=\zcluster{m0329}$), which were considered as the blind test set. Such experiment is particularly suitable to evaluate the model capability to predict the cluster membership of sources extracted from clusters unused during training. Furthermore, in this experiment, we varied the training configuration based on three redshift ranges centered on \A{370} cluster redshift (named as \textit{narrow}, \textit{intermediate} and \textit{large}, see Sect.~\ref{ss:exp2} and Fig.~\ref{fig:exp2:layout}), exploiting the \textit{mixed*} band configuration.  This experiment has been described in Sect.~\ref{ss:exp2}. Table~\ref{tab:exp2} outlines the results achieved on the three configurations together with Fig.~\ref{fig:exp2}.
As second step, we analysed the CNN classification capabilities by separating, respectively, brighter from fainter (\textit{EXP2a}), and redder from bluer objects (\textit{EXP2b}). Concerning the magnitude threshold, we split the \cm{} $F814$ distribution into two equal-sized sets ($F814$ limits are $22.0mag$, $21.7mag$, and $21.6mag$ for, respectively, \A{370}, \M{0329} and \MS{2137}). Regarding the colour split, we exploited the correlation between the Balmer break and the differential colour, as shown in \cite{girardi2015}: $(F814 - F160)_{\text{diff}} = (F814 - F160W_{\text{obs}} - \text{CM}(F814)$, that is, the difference between the observed colour and the one of the colour-magnitude (CM) relation at a given magnitude. For each cluster, we fitted the CM sequence using a robust linear regression \citep{cappellari2013} involving spectroscopic confirmed members. By applying this kind of correction to the source colour, redder members were centered around zero, while bluer objects have differential colours around $-0.2$mag. The differential colour thresholds were $-0.160$, $-0.165$, $-0.157$ mag for, respectively, \A{370}, \M{0329} and \MS{2137}. For both experiments, we opted for a \textit{large} ensemble and \textit{mixed*} band configuration. The results are shown in Table~\ref{tab:exp2:magcolsplit}.

\begin{table}[htbp]\caption{Percentage performances on a blind test set related to the \textit{EXP2} experiment.}\label{tab:exp2}\centering
\begin{tabular}{clccc}
\hline
\multicolumn{2}{l}{stacked}& \textit{Narrow} & \textit{Intermediate} & \textit{Large} \\
& AE & $84.5\pm0.6$ & $85.5\pm0.4$ & $\mathbf{86.6\pm0.3}$ \\
\cm{} & pur & $79.6\pm1.2$ & $\mathbf{83.2\pm0.2}$ & $82.5\pm0.6$ \\
\% & comp & $87.6\pm0.8$ & $83.9\pm0.8$ & $\mathbf{88.5\pm0.4}$ \\
& F1 & $83.3\pm1.2$ & $83.6\pm0.2$ & $\mathbf{85.4\pm0.6}$ \\\hline

\multicolumn{2}{l}{\A{370}} &\textit{Narrow} & \textit{Intermediate} & \textit{Large} \\
& AE & $85.4\pm0.7$ & $86.6\pm0.3$ & $\mathbf{87.4\pm0.3}$ \\
\cm{} & pur & $80.3\pm1.4$ & $84.5\pm0.2$ & $\mathbf{83.9\pm0.7}$ \\
\% & comp & $86.5\pm0.7$ & $83.1\pm0.6$ & $\mathbf{86.6\pm0.6}$ \\
& F1 & $83.3\pm1.4$ & $83.8\pm0.2$ & $\mathbf{85.1\pm0.7}$ \\\hline

\multicolumn{2}{l}{\M{0329}} &\textit{Narrow} & \textit{Intermediate} & \textit{Large} \\
& AE & $81.7\pm0.5$ & $83.5\pm0.5$ & $\mathbf{84.8\pm0.3}$ \\
\cm{} & pur & $76.9\pm0.7$ & $\mathbf{79.2\pm0.5}$ & $\mathbf{79.2\pm0.4}$ \\
\% & comp & $90.0\pm0.6$ & $90.4\pm0.4$ & $\mathbf{93.9\pm0.4}$ \\
& F1 & $82.9\pm0.7$ & $84.4\pm0.5$ & $\mathbf{85.9\pm0.4}$ \\\hline

\multicolumn{2}{l}{\MS{2137}} &\textit{Narrow} & \textit{Intermediate} & \textit{Large}\\
& AE & $84.1\pm1.1$ & $82.4\pm2.3$ & $\mathbf{85.4\pm0.7}$ \\
\cm{} & pur & $81.5\pm1.9$ & $\mathbf{84.4\pm1.8}$ & $82.3\pm1.0$ \\
\% & comp & $87.6\pm1.5$ & $77.1\pm4.8$ & $\mathbf{88.9\pm0.7}$ \\
& F1 & $84.2\pm1.9$ & $80.0\pm1.8$ & $\mathbf{85.4\pm1.0}$ 
\end{tabular}
\tablefoot{Performances have been split between the three test clusters: \A{370} ($z=\zcluster{a370}$), \MS{2137} ($z=\zcluster{m2137}$), \M{0329} ($z=\zcluster{m0329}$) and their stacking. Best results are emphasised in bold. For ease of reading, only statistics related to the \cm{} class are reported, together with the average efficiency (AE), which refers to both classes.}
\end{table}

\subsection{EXP3}\label{app:Exp3}
\begin{figure*}[htbp]
   \centering
   \includegraphics[width=2\columnwidth]{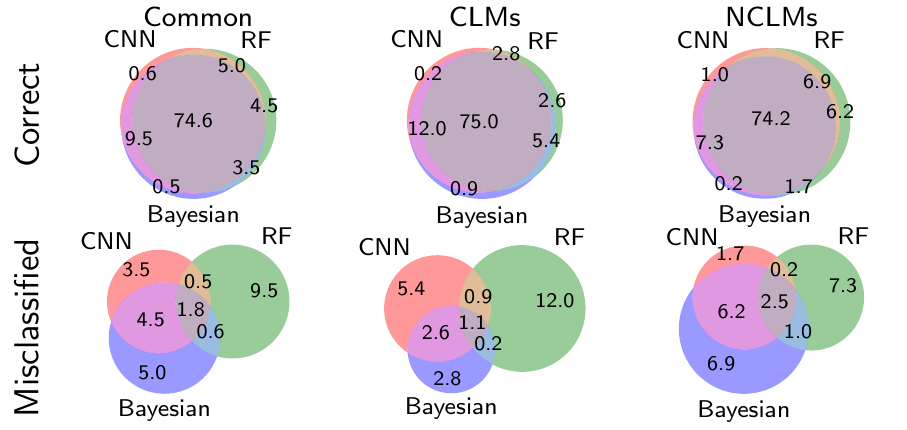}
   \caption{Venn diagrams reporting the percentages of membership predictions performed by three different methods (CNN, RF, and Bayesian), measured on the common blind test set, obtained by combining the four clusters \R{2248}, \M{0416}, \M{1206}, and \M{1149} (\textit{EXP3}). On the columns, the common areas refer to the available shared sources, respectively, $460$ \cm{s} and $519$ \ncm{s}). On the rows, common predictions are split between correct and incorrect classifications. Global commonalities can be derived by summing values on the rows.} \label{fig:exp3}
\end{figure*}

\begin{figure*}[htbp]
   \centering
   \includegraphics[width=2\columnwidth]{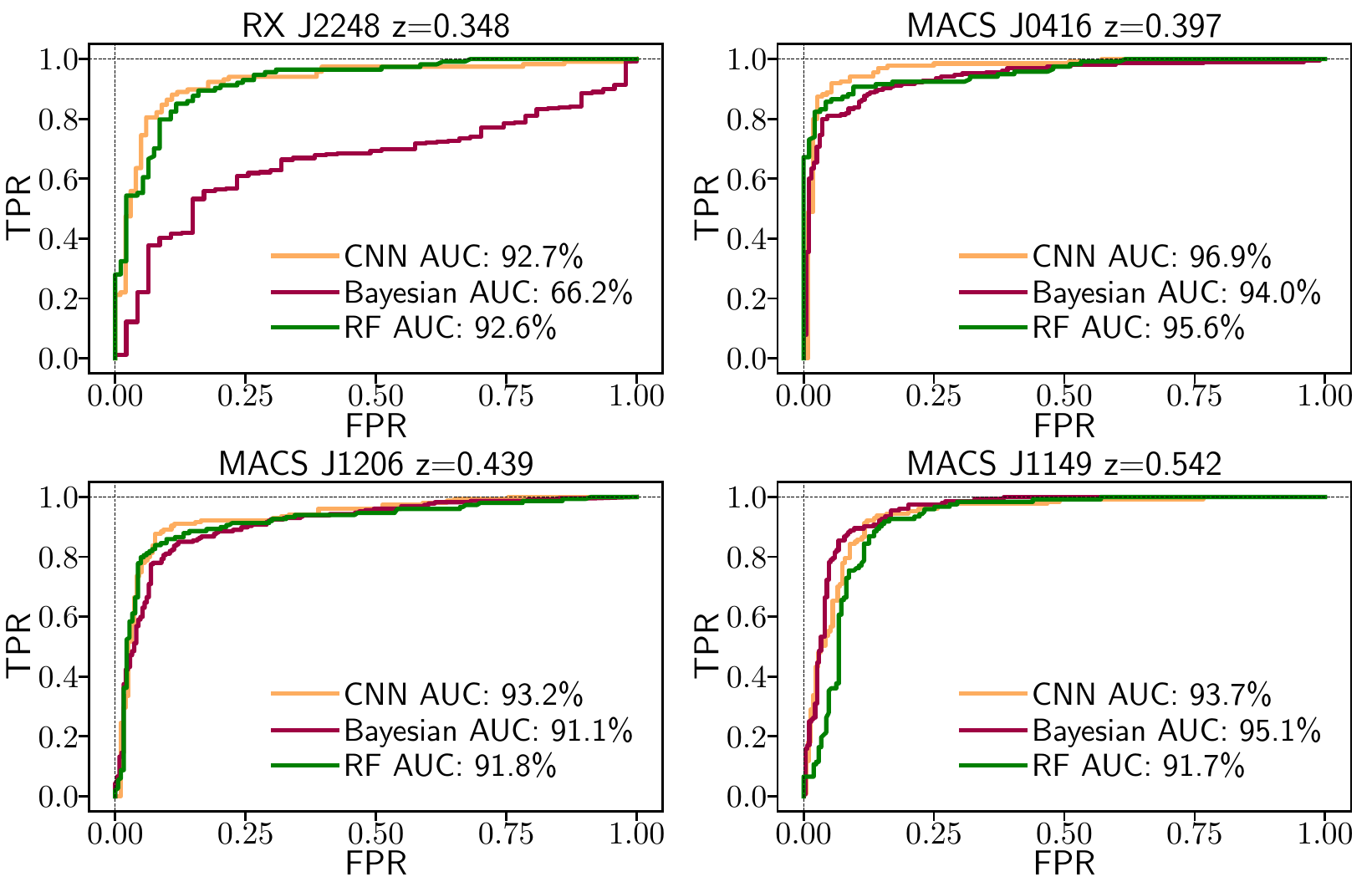}
   \caption{Comparison between the image-based CNN ant two photometric catalogue-based approaches, RF and Bayesian method (\textit{EXP3}) in term of ROC curves for the four clusters: \R{2248} (top-left panel), \M{0416} (top-right panel), \M{1206} (bottom left panel), \M{1149} (bottom right panel).}\label{fig:exp3:roc:clusters}
\end{figure*}

This test was devoted to the comparison of CNN performance with two different photometry-based methods, exploiting a random forest classifier \citep{breiman:2001} and a Bayesian model \citep{grillo2015}. Both techniques critically use multi-band photometric information, for example, magnitudes and colours. This experiment has been outlined in Sect.~\ref{ss:exp3}. The Bayesian method has already been applied in order to enlarge the cluster member selection, including galaxies without spectroscopic information, for four clusters: \R{2248}, \M{0416}, \M{1206}, and \M{1149} \citep{grillo2015, caminha2016, caminha2017b, treu2016}. We compare these methods with our CNN, trained with the \textit{mixed*} band configuration, constraining the results to these four involved clusters. The comparison is summarised in Table~\ref{tab:exp3} in term of statistical estimators, whereas, in Fig.~\ref{fig:exp3:roc} and Fig.~\ref{fig:exp3:roc:clusters}, it is shown in terms of ROC curves (see Sect.~\ref{ss:estimators}), in Fig.~\ref{fig:exp3} in terms of commonalities among predictions. Particularly, we also compared performances between CNN and photometric methods by computing the differences: $\Delta_{estim}=estim_{CNN}-max\{estim_{RF},\, estim_{Bayesian}\}$ for $estim \in [pur,\, compl,\, F1,\, AE]$, that is, the difference between CNN metrics and the corresponding maximum scores achieved by RF or Bayesian model. All these differences are listed in the last column of Table~\ref{tab:exp3}, together with the average among these $\Delta$s for each cluster (rows $\mu_\Delta$). 

An additional comparison of the three methods based on common membership predictions (see Fig.~\ref{fig:exp3}), is discussed in Sect.~\ref{ss:exp3}. 

\begin{table}[htbp]\caption{Comparison between our image-based CNN model and two different photometric catalogue-based approaches, referred to the \textit{EXP3} experiment.}\label{tab:exp3}\centering
\begin{tabular}{llcccc}\\\hline\hline
&\multicolumn{5}{c}{\R{2248} $z=\zcluster{r2248}$} \\
&& CNN & RF & Bayesian & $\Delta$ \\\hline
& AE & \textbf{88.1} & 86.5 & 85.9 & 1.6 \\
& pur & \textbf{88.3} & 87.7 & 80.9 & 0.6 \\
\cm{} & compl & 89.8 & 87.7 & \textbf{96.1} & -6.3 \\
& F1 & \textbf{89.1} & 87.7 & 87.8 & 1.3 \\
& pur & 87.9 & 85.1 & \textbf{94.4} & -6.5 \\
\ncm{} & compl & \textbf{86.1} & 85.1 & 74.4 & 1.0 \\
& F1 & \textbf{87.0} & 85.1 & 83.2 & 1.9 \\\hline
&$\mu_\Delta$ &\multicolumn{4}{c}{$-0.91\pm1.42$}\\\hline\hline

&\multicolumn{5}{c}{\M{0416} $z=\zcluster{m0416}$} \\
&& CNN & RF & Bayesian & $\Delta$ \\\hline
& AE & \textbf{92.2} & 89.2 & 87.1 & 3.0 \\
& pur & \textbf{93.3} & 93.0 & 84.6 & 0.3 \\
\cm{} & compl & 87.1 & 86.5 & \textbf{91.2} & -4.1 \\
& F1 & \textbf{91.5} & 89.7 & 87.8 & 1.8 \\
& pur & 89.0 & 84.5 & \textbf{90.0} & -1.0 \\
\ncm{} & compl & \textbf{96.9} & 92.3 & 82.7 & 4.6 \\
& F1 & \textbf{91.5} & 88.3 & 86.2 & 3.2 \\\hline
&$\mu_\Delta$ &\multicolumn{4}{c}{$1.11\pm1.12$}\\\hline\hline

& \multicolumn{4}{c}{\M{1206} $z=\zcluster{m1206}$} \\
&& CNN & RF & Bayesian & $\Delta$ \\\hline
& AE & \textbf{89.7} & 87.9 & 85.0 & 1.8 \\
& pur & 89.9 & \textbf{90.4} & 80.2 & -0.5 \\
\cm{} & compl & 86.5 & 81.9 & \textbf{91.2} & -4.7 \\
& F1 & \textbf{88.2} & 85.9 & 85.3 & 2.3 \\
& pur & 89.6 & 86.3 & \textbf{90.8} & -1.2 \\
\ncm{} & compl & 92.3 & \textbf{92.9} & 79.4 & -0.6 \\
& F1 & \textbf{90.9} & 89.7 & 84.7 & 1.2 \\\hline
&$\mu_\Delta$ &\multicolumn{4}{c}{$-0.24\pm0.90$}\\\hline\hline

& \multicolumn{5}{c}{\M{1149} $z=\zcluster{m1149}$}\\
&& CNN & RF & Bayesian & $\Delta$ \\\hline
& AE & \textbf{89.4} & 86.9 & 85.5 & 2.5\\
& pur & \textbf{82.3} & 78.8 & 71.8 & 3.5\\
\cm{} & compl & 91.3 & 88.5 & \textbf{98.0} & -6.7\\
& F1 & \textbf{86.6} & 83.4  &82.9 & 3.2\\
& pur & 94.5 & 92.7 & \textbf{98.6} & -4.1\\
\ncm{} & compl & \textbf{88.3} & 86.0 & 78.4 & 2.3\\
& F1 & \textbf{91.3} & 83.4 & 87.4 & 3.9\\\hline
&$\mu_\Delta$ &\multicolumn{4}{c}{$0.66\pm1.60$}\\\hline\hline
\end{tabular}
\tablefoot{The comparison involves two different model: a Random Forest and a Bayesian method, applied on photometric tabular information of four clusters: \R{2248} ($z=\zcluster{r2248}$), \M{0416} ($z=\zcluster{m0416}$), \M{1206} ($z=\zcluster{m1206}$) and \M{1149} ($z=\zcluster{m1149}$). Last column ($\Delta$) shows the difference between CNN estimators and the best between the two photometric approaches, i.e. $\Delta_{estim} = estim_{CNN}-max\{estim_{RF},\, estim_{Bayesian}\}$ for $estim \in [pur,\, compl,\, F1,\, AE]$, while rows $\mu_\Delta$ list the averages among these $\Delta$s for each cluster.}
\end{table}

\end{document}